\documentclass[traditabstract]{aa} % for the abstract without structuration 
\usepackage{graphicx}
\usepackage{amssymb}
\usepackage{amsmath}
\usepackage{natbib}
\usepackage[flushleft]{threeparttable}
\usepackage{longtable}
\usepackage{supertabular}
\usepackage{multirow}

\usepackage{txfonts}
\begin{document}

   \title{Isolated ellipticals and their globular cluster systems I:
     Washington photometry of NGC 3585 and NGC 5812}

   \subtitle{}

   \author{Richard R. Lane\inst{1}\fnmsep\thanks{rlane@astro-udec.cl},
          Ricardo Salinas\inst{2} \and Tom Richtler\inst{1}}

   \institute{Departamento de Astronom\'ia Universidad de Concepci\'on,
Casilla 160 C, Concepci\'on, Chile
\and
Finnish Centre for Astronomy with ESO, University of Turku, V\"ais\"al\"antie
20, FI-21500 Piikki\"o, Finland}

   \date{Received .....; accepted .....}

% \abstract{}{}{}{}{} 
% 5 {} token are mandatory
 
  \abstract
  % {} leave it empty if necessary {}
  {The globular cluster (GC) systems of isolated elliptical galaxies have only
    recently begun to be studied in detail, and may exhibit morphological
    connections to the evolutionary histories of their hosts. Here we present
    the first in a series of wide-field analyses of the GC systems of the
    isolated ellipticals -- Washington $C$ and $R$ photometry of NGC 3585 and
    NGC 5812 down to $R\sim24$\,mag. The GC systems are characterised, with
    each system displaying both the ``Universal'' blue peak at $(C-R)\sim1.3$,
    and a red peak, but each with differing strengths. The total number of GCs
    in each system, and their specific frequencies, are estimated. The GC
    colours and specific frequencies are highly indicative that the host
    galaxy environment plays a role in shaping its GC system. We produce, and
    subtract, accurate models of each galaxy, revealing interesting underlying
    features, including the first definitive evidence that NGC 5812 is
    interacting with a dwarf companion galaxy. From the galaxy models we also
    determine surface brightness and colour profiles. Both colour profiles
    appear quite flat and with $(C-R)\sim1.7$ and we discuss the apparent
    youth of NGC 3585 in the context of this work.}
   { }
   { }
   { }


   \authorrunning{Lane, Salinas \& Richtler}
   \titlerunning{The GC systems of NGC 3585 and NGC 5812}

   \maketitle
%
%________________________________________________________________

\section{Introduction}

There is some strong evidence that giant elliptical galaxies grow their
extended stellar haloes slowly, through accretion, around a dense, compact
core \cite[e.g.][]{vanDokkum10,vanDokkum12}. If this is the case, their halo
globular cluster (GC) populations must have been built up in the same
manner. Since most elliptical galaxies reside in clusters or groups, where
significant quantities of material is available for accretion, the question
naturally arises -- what about in isolated ellipticals which reside in very
low density environments?  Interestingly, clues to this may be gleaned from
their GC populations.

Recent work by \cite{Tal12} indicates that many, possibly all, isolated
elliptical (IE) galaxies are the result of late mergers and are, therefore,
not ancient relics. Currently very little is known about the GC populations of
IEs. To date, there are only three giant IEs that have had their GC systems
studied in any detail, those of NGC 720 \cite[][]{KisslerPatig96}, NGC 821
\cite[][]{Spitler08}, and NGC 3585 \cite[][]{Hempel07,Humphrey09}
(\citealt{Cho12} also recently added a photometric investigation of NGC 3818
-- the only one of the 10 galaxies studied classified as an elliptical and
with an environmental density $\lesssim0.2$\,Mpc$^{-3}$). None of the galaxies
were found to have rich GC systems. From those studies, the estimated total
numbers of GCs, and their specific frequencies, for NGC 720, NGC 821, NGC 3585
and NGC 3818, respectively, are ${\rm GC_{tot}}\sim660$ \& $S_N\sim2.2$, ${\rm
  GC_{tot}}\sim320$ \& $S_N\sim1.32$, ${\rm GC_{tot}}\sim90$ \& $S_N\sim0.47$
and ${\rm GC_{tot}}\sim240$ \& $S_N\sim1.36$.

Here we report on two interesting examples of isolated elliptical galaxies,
namely NGC 3585 and NGC 5812. NGC 3585 is classified as E6 by
\cite{deVaucouleurs91} and a high rotational velocity of at least
\cite[157$\pm$3\,km\,s$^{-1}$ --][]{Koprolin00}, or even higher
\cite[e.g. 206$\pm$3\,km\,s$^{-1}$ --][]{Terlevich02}. Furthermore, it appears
to have some asymmetry in its outer isophotes \cite[][]{Tal09}. Both of these
properties are indicative of a recent merger, however, remnants of a merger
have yet to be uncovered. In contrast, NGC 5812 appears to be very spherical
(classified as E0 by \citealt{deVaucouleurs91} and has an ellipticity of only
0.05, e.g. \citealt{Bender88}). Note that these galaxies are considered
isolated in that they have no companion galaxies within 75\,kpc, although NGC
5812 may have a {\it small dwarf} companion \cite[][also see
  Section~\ref{5812dwarf} for a discussion of the companion of NGC
  5812]{Madore04,Tal09}. Table~\ref{paramtable} summarises the generic
properties of each galaxy. The age quoted in Table~\ref{paramtable} for NGC
3585 is the time since the most recent star formation event, derived from six
colours and four Lick line indices \cite[see][for details]{Michard06}. Another
age estimate exists in the literature \cite[$\sim3.1$\,Gyr --][]{Terlevich02},
however, it is unclear how the authors of the study arrived at this value. For
NGC 5812 the age is estimated using $<$Fe$>$ and the line strengths of
H$\beta$ and Mg$b$ in accordance with single-burst stellar population models
\cite[see][for details]{Trager00}. The distances quoted in Table
\ref{paramtable} are the mean distances from various studies.

\begin{table*}[!ht]
\centering
\begin{threeparttable}
\caption{Literature galaxy parameters}\label{paramtable}
\begin{tabular}{@{}cccccccc@{}}
\hline
\hline
Name & Age (Gyr) & [Fe/H]$^{\rm a}$ & $(V-R)$ & Extinction (E[B-V])$^{\rm b}$
& M$_V^{\rm c}$ & Distance (Mpc) & ${\rm pc}/''$\\
\hline
NGC 3585 & 1.7$^{\rm a}$ & $>0.5$ & 0.59$^{\rm d}$ & 0.057 & -21.8 ($<380''$) & 18.3$^{\rm e}$ & 89\\
NGC 5812 & 5.9$^{\rm f}$ &   0.39 & 0.66$^{\rm g}$ & 0.077 & -21.4 ($<270''$) & 27.7$^{\rm e}$ & 134\\
\hline
\end{tabular}
\begin{tablenotes}
\item[a] \cite{Michard06}
\item[b] \cite{Schlegel98}; \cite{Schlafly11}
\item[c] This work (Section~\ref{galmodsec})
\item[d] \cite{Prugniel98}
\item[e] NASA/IPAC Extragalactic Database (NED) -- {\tt http://ned.ipac.caltech.edu}
\item[f] \cite{Trager00}
\item[g] \cite{Ferrari99}
\end{tablenotes}
\end{threeparttable}
\end{table*}

Presented here are deep, wide field Washington $C$ and Harris $R$
photometry\footnote{Note that the spectral response of the Harris $R$ and
  Washington $T_1$ filters are effectively equivalent for objects with
  $(C-T_1)\lesssim3.5$, however the $R$ filter is preferable due to its higher
  throughput \cite[][]{Geisler96}.} of NGC 3585 and NGC 5812 with the aim of
characterising their GC systems, as well as the surface brightness and colour
profiles of the galaxies themselves. The Washington photometric system
\cite[][]{Canterna76} has been chosen as it has the advantage of being a good
discriminator between compact blue background galaxies and GC candidates
\cite[][]{Dirsch03a}. Furthermore, an apparently Universal peak exists in the
$(C-R)$ colour of old globular cluster populations associated with elliptical
galaxies \cite[e.g.][and references therein]{Richtler12}, which warrants
further investigation.

%This is especially true in the Washington photometric system
%\cite[][]{Canterna76}, in part because the Washington system provides very
%good metallicity discrimination \cite[e.g.][]{Friel90}; the Washington
%$(C-T_1)$ colour has at least twice the senstivity to metallicity as the
%$(V-I)$ colour \cite[e.g.][]{Geisler99}. This metallicity sensitivity removes
%the necessity for combining infrared and optical datasets for this type of
%study \cite[as used by][for example]{Hempel07}.

In the majority of elliptical galaxies, the $(C-T_1)$ colour profiles of the
GC systems are bimodal, however, the GC systems of {\it truly} isolated
elliptical galaxies may be unimodal, depending on the assumed formation
scenario \cite[e.g.][]{Brodie00,Lee08}.  Indeed, in many cases it seems that
red (metal rich) GC populations may have formed {\it in situ} along with the
galaxy, while the bluer (more metal poor) GCs arrived later as part of the
hierarchical merger process, assuming mainly minor mergers
\cite[][]{Lee08,Elmegreen12}. Moreover, if an early-type galaxy forms from the
merger of two gas rich components, an extended burst of GC formation can take
place early on. In this case the colour distribution of the GCs is likely to
be much broader with less obvious bimodality.  Observations of GC systems of
elliptical galaxies in Washington filters are, therefore, very useful in
tracing of the formation and merger histories of their parent galaxy. In
addition, the $(C-T_1)$ colour profile of the galaxy itself is strongly
indicative of its recent history. For example ``wet'' (gas rich) major mergers
often trigger star bursts \cite[e.g.][and references therein]{ElicheMoral10},
leading to variations in the colour profile.

In Section~\ref{datasection} we describe the data aquisition, reduction and
photometric calibration. In Section~\ref{galmodsec} the elliptical galaxy
models and point source catalogues are presented, as well as a discussion of
the apparent dwarf companion of NGC 5812. In Section~\ref{SBsection} we
analyse the galactic surface brightnesses, and in Section~\ref{coloursection}
their colour profiles, based on these models. We present our analyses of the
galactic GC populations, based on our photometry, in
Section~\ref{GCsection}. Finally, in Section~\ref{conclusions} we discuss our
findings and present our conclusions. Note that the tangential scales used in
this paper, at the distances to NGC 3585 and NGC 5812, are $\sim89$ and
$\sim134$ parsecs per arcsecond, respectively (see Table~\ref{paramtable}).

\section{Data}\label{datasection}

\subsection{Aquisition and reduction}

The data were obtained using the MOSAICII camera on the CTIO 4m Blanco
telescope on Cerro Tololo, Chile, with the Washington $C$ (c6006) and Harris
$R$ (c6004) filters, under photometric conditions, on March 19-20, 2010. The
field of view for the camera is $36'\times36'$, with a pixel scale of
$0.27''$. For each of the two fields we performed five dithered exposures of
1440\,s and one of 300\,s in the $C$ filter and five dithered exposures of
720\,s plus one of 60\,s in the $R$ filter. The mean seeing on the two nights
was $1.1''$ and $1.3''$, respectively. Several Washington standard star fields
by \cite{Geisler90,Geisler96} were also observed for calibration purposes.

The removal of the instrumental signatures, astrometric alignment, and
coaddition of the individual images was carried out using the THELI data
reduction pipeline \cite[][]{Erben05}. Sky subtraction was also performed with
THELI; we found that the ideal method came from taking the median value of
each chip as the sky value used for subtraction. This also worked for those
chips containing the galaxy light because, in general, the galaxy light
extended across at least two chips and, therefore, covered much less than half
of each chip. Sky subtraction was performed as it is imperative for producing
reliable colour maps (Section \ref{coloursection}).

\subsection{Photometry}

Due to a lack of airmass variation during our observations it was not possible
to determine accurate values for the extinction coefficients for our
photometric zeropoint calibration. To determine our photometric zeropoint,
therefore, we have taken extinction coefficients, and colour terms, from the
literature, both of which show very little variation over time
\cite[e.g. see][]{Dirsch03a,Dirsch03b,Harris07}. We have adopted for the $C$
filter a colour term $A_{C_{(C-R)}}=0.1$ and extinction coefficient of
$A_{(X,C)}=0.3$ and for the $R$ filter $A_{R_{(C-R)}}=0.078$ and
$A_{(X,R)}=0.030$.

The photometric zeropoint calibrations were then performed using IRAF routines,
as follows. Instrumental magnitudes of the standard stars were calculated
using the aperture photometric {\it apphot} package, ensuring our aperture
sizes matched those used by the \cite{Geisler90,Geisler96} studies. Our
photometric zeropoint was then calculated for each standard star using:

\begin{eqnarray}
C_{\rm i} = {\rm ZP}_C + C_{\rm std} + (C-R)_{\rm std} \times A_{C_{(C-R)}} -
A_{(X,C)} \times X_{\rm std_{\it C}}\\
R_{\rm i} = {\rm ZP}_R + R_{\rm std} + (C-R)_{\rm std} \times A_{R_{(C-R)}} -
A_{(X,R)} \times X_{\rm std_{\it R}}
\end{eqnarray}
where $C_{\rm i}$ and $R_{\rm i}$ are the instrumental magnitudes, ZP$_C$ and ZP$_R$ are the zeropoints being calculated, $(C-R)_{\rm std}$
is the $(C-R)$ colour of the standard star from \cite{Geisler90,Geisler96},
$A_{C_{(C-R)}}$ and $A_{R_{(C-R)}}$ are the colour terms, $A_{(X,R)}$ and
$A_{(X,C)}$ are the extinction coefficients and $X_{\rm std_{\it C}}$ and
$X_{\rm std_{\it R}}$ are the airmasses. The mean zeropoint for all $\sim50$
standard stars in the $C$ filter was $0.26\pm0.09$ and for the $R$ filter
$0.73\pm0.07$ with the stated uncertainties being the standard deviation of
the calculated zeropoints.

The {\it daophot} package {\it psf} was then used to determine the psf
magnitudes of each star in the galaxy-subtracted (Section~\ref{galmodsec})
science fields. Finally, before zeropoint correction, we sanity-checked our
psf magnitudes by applying {\it apphot} to several isolated stars in each
field, with aperture radii matching the psf radii used by the {\it psf}
routine. The mean difference between our psf and aperture photometry for these
15 stars was $<0.02$ mags in both filters, well within the zeropoint
uncertainties.

\section{Galaxy models and point source evaluation}\label{galmodsec}

To help facilitate our search for point sources within our fields, and to
determine the surface brightness profiles of the galaxies, we subtracted the
galaxy light using the IRAF packages {\it ellipse} and {\it bmodel}. Before
producing the {\it ellipse} model, it was necessary to replace the central,
overexposed, region from each galaxy with the central region from the short
exposure in each filter, to ensure the best possible model at all radii. The
original $R$ images, with central regions replaced as described, and those
with the $ellipse$ model subtracted can be seen in Figures
\ref{NGC3585galaxyfig} \& \ref{NGC5812galaxyfig}. A slightly noisy, small,
rectangular central region can be seen, particularly in the lower panel of
Figure~\ref{NGC5812galaxyfig}, which is the region that was replaced with that
of the short exposure.

\subsection{NGC 3585}

The {\it ellipse} model was applied allowing the centre of the isophotes to
shift between successive iterations. Linear steps between iterations were not
enforced and we allowed a maximum semi-major axis of 1400 pixels ($\sim380''$)
to ensure that the model did not extend to unreasonably large radii. In order
for the {\it ellipse} task to ignore bright sources we set the {\it usclip}
parameter to 5, enforcing a 5 sigma clipping criterion to deviant points above
the mean, which we found to be optimum. To ensure that our sigma clipping
method produced reliable {\it ellipse} models, we performed two tests. First
we used a pixel mask based on the CHECKIMAGE\_TYPE variable ``segmentation''
output of Sextractor and second we employed the {\it ellipse} task in
interactive mode, which allows for interactive masking during the modelling
procedure. The differences between the three models were negligible.

\begin{figure}
  \begin{centering}
  \includegraphics[width=0.48\textwidth]{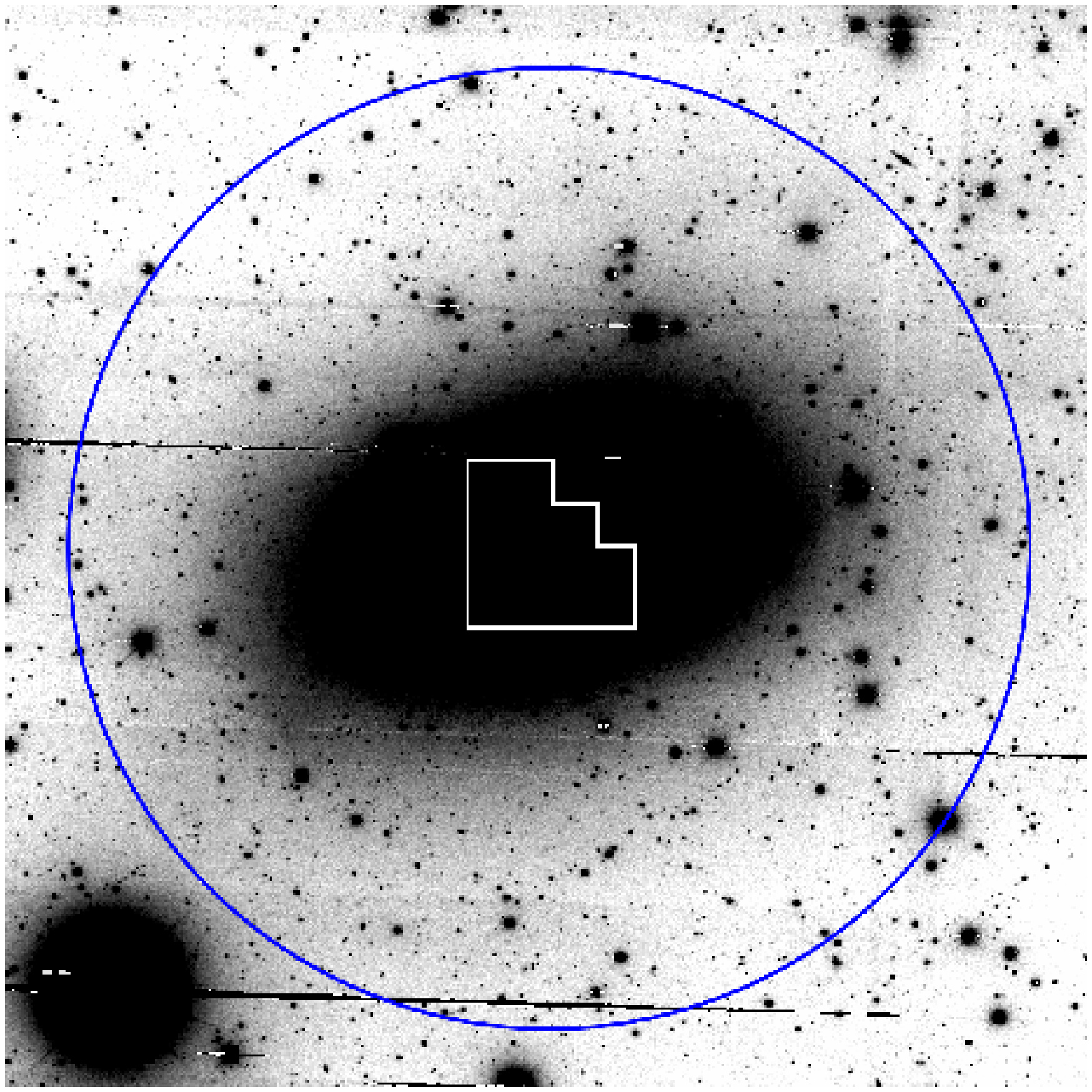}
  \includegraphics[width=0.48\textwidth]{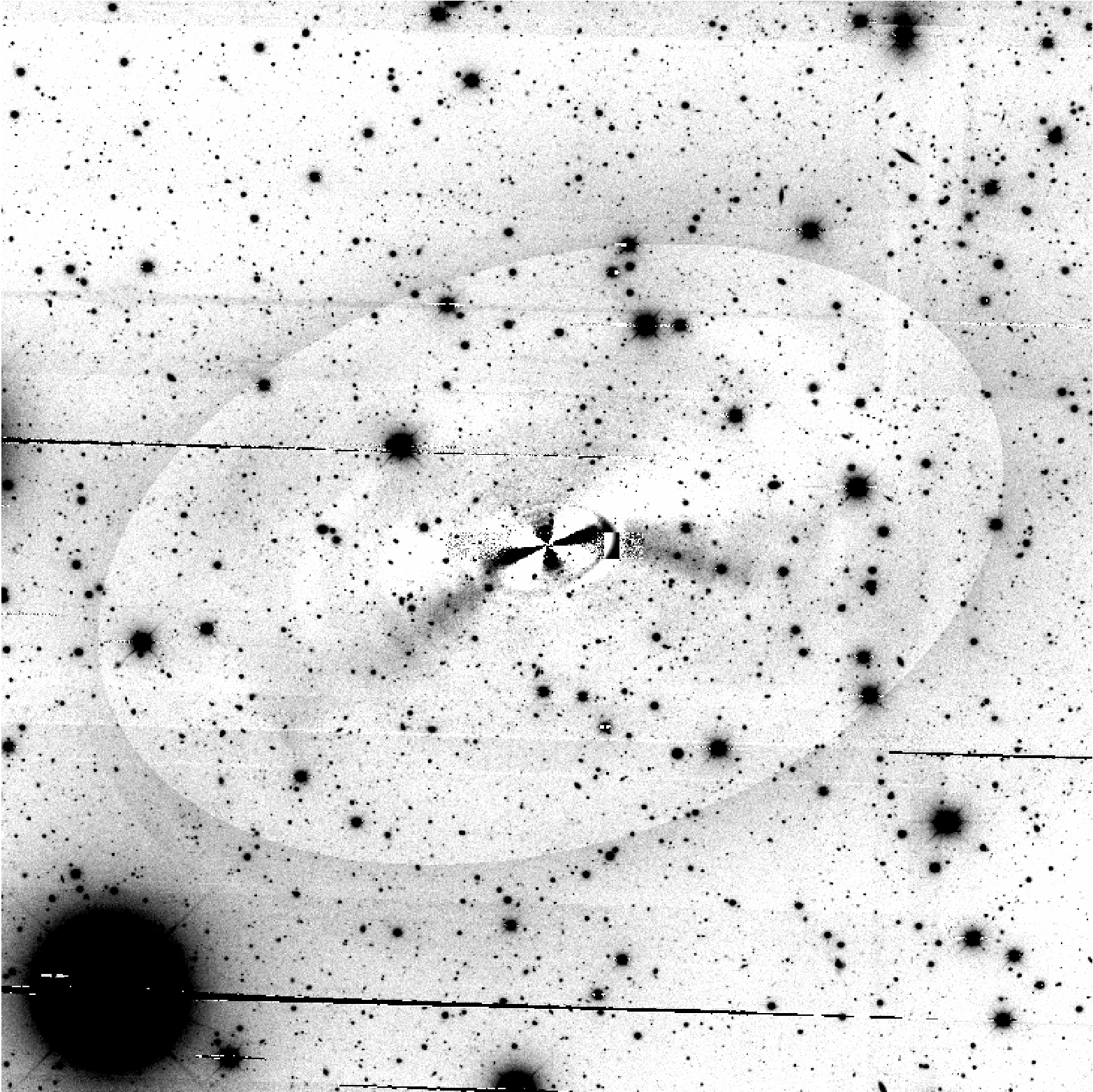}
  \caption{$R$ images of NGC 3585 The upper panel shows the original image
    while the lower panel shows the image after the galaxy light is removed
    with {\it ellipse}. The images are $\sim15'\times15'$. North is up and
    East is to the left. In the upper panel, the approximate size of the WFPC2
    footprint (white polygon), and the approximate radius beyond which our
    point source counts drop to the background level (blue circle), are
    included for comparison (see Section \ref{GCsection}).}
  \label{NGC3585galaxyfig}
  \end{centering}
\end{figure}

Note that in the lower panel of Figure~\ref{NGC3585galaxyfig} some small
residual galaxy light remains beyond the edge of the subtracted model. Due to
the small fluxes in these regions the modelling did not perform well outside
1400 pixels. Note, however, that the difference between the flux for the
regions just inside and just outside the edge of the model are very small
($\lesssim0.01$ counts, normalised to a 1 second exposure).

The {\it ellipse} task determines the total enclosed flux of the galaxy light
within each isophote. Based on the final isophote (with a semi-major axis of
$380''$), which encloses all the available galaxy light, we calculate the
apparent magnitude of NGC 3585 to be m$_R\sim8.9$. Assuming
$(V-R)\sim0.59\pm0.01$ (Table~\ref{paramtable}) we find a $V$-band apparent
magnitude of m$_V\sim9.5$ (within the final isophote). We also calculate, from
the {\it ellipse} model output, an effective radius of $R_e\sim80''$ with an
enclosed surface brightness of $R\sim6.7$ mag per square arcsec. We take the
effective radius as the radius which encloses half the flux from the galaxy,
directly from the {\it ellipse} model output tables.

\cite{Tal09} showed that NGC 3585 has some asymmetry in its outer isophotes
and derive an interaction parameter of 0.048, indicating a lack of any clear
interaction signature. Our {\it ellipse} model shows a small drop in
ellipticity from the inner to the outer isophotes (from $\sim0.42$ to
$\sim0.37$), although the position angle of the isophotes remains relatively
constant at $\sim285^\circ$, whereas the A4 parameter drops from $\sim0.1$ in
the inner regions to $\sim0.001$ for the outer isophotes. The A4 parameter is
a measure of how ``discy'' or ``boxy'' an isophote is \cite[see][for a
  detailed explanation of the A4 parameter]{Bender88}, with positive values
indicating discy isophotes and negative values indicating boxy isophotes. We
do not see anything unusual in this slight shift to more circular isophotes
at large radii, or the drop in the A4 parameter, as both these effects are
expected for a galaxy with a discy core.

After subtracting our model, a lot of small-scale structure is
revealed. Despite the small interaction parameter assigned by Tal et al. the
variety of visible structure leaves little doubt that merger activity has left
its imprint. The inner disc is very prominent and there are also apparent
faint shells and ripples. Most noteworthy are the two bright residuals
emanating from the disc, to the South East and West-South West, which appear
to give the impression of a $\cap$-shaped structure. However, the apparent
symmetry may be a coincidence. We have no reason to doubt the veracity of our
galaxy model (see Section~\ref{5812galmod}), so these visible substructures
are indicative of a complex dynamical system.

\subsection{NGC 5812}\label{5812galmod}

The {\it ellipse} model was applied in the same manner as for NGC 3585,
however we restricted the maximum semi-major axis to 1000 pixels
($\sim270''$), to ensure that the model did not extend to unreasonable
radii. Again we set the {\it usclip} parameter to 5, which we found to be
optimum for removing bright sources from the final model (we performed the
same tests as for NGC 3585 to ensure our models were reliable). The ellipse
model for NGC 5812 was compared to the calculated isophotal measurements by
\cite{Bender88} and found very good agreement with those authors for all
parameters, e.g. position angle, ellipticity, A3 and A4. \cite{Bender88} did
not analyse NGC 3585, however, due to the nearly identical match between our
derived parameters and those of that paper for NGC 5812, we see no reason to
doubt the veracity of our model for NGC 3585.

From the flux enclosed by the final isophote (with a semi-major axis of
$\sim270''$), we find an apparent $R$ magnitude of m$_R\sim10.1$, derived from
the flux internal to the outer isophote from the {\it ellipse} task. This
relates to an apparent $V$ mag of m$_V=10.8$, since $(V-R)\sim0.66$
(Table~\ref{paramtable}). The effective radius, determined in the same way as
for NGC 3585, is $35''$ with an enclosed surface brightness of $\sim8.0$ mag
per square arcsec.

\begin{figure}
  \begin{centering}
  \includegraphics[width=0.48\textwidth]{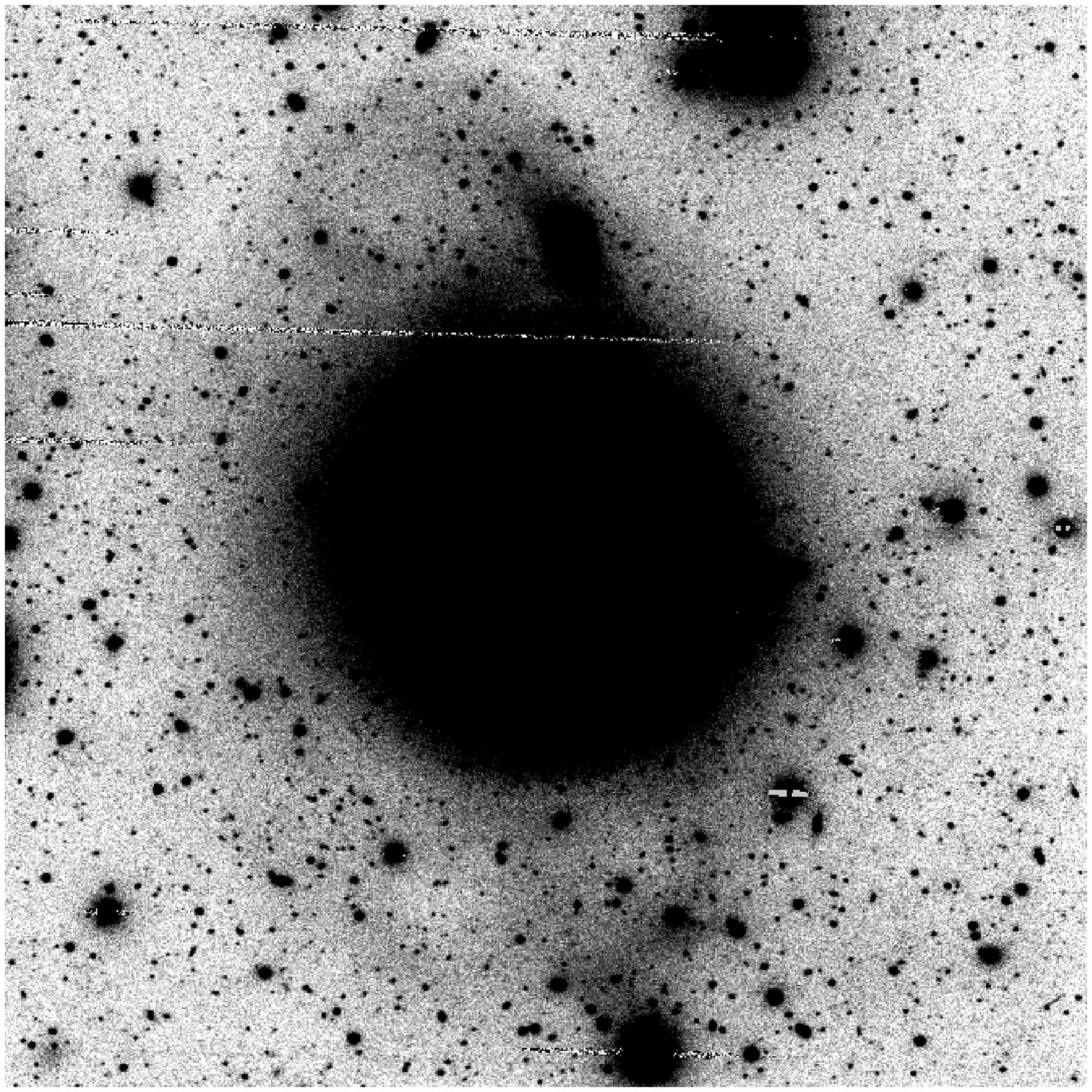}
  \includegraphics[width=0.48\textwidth]{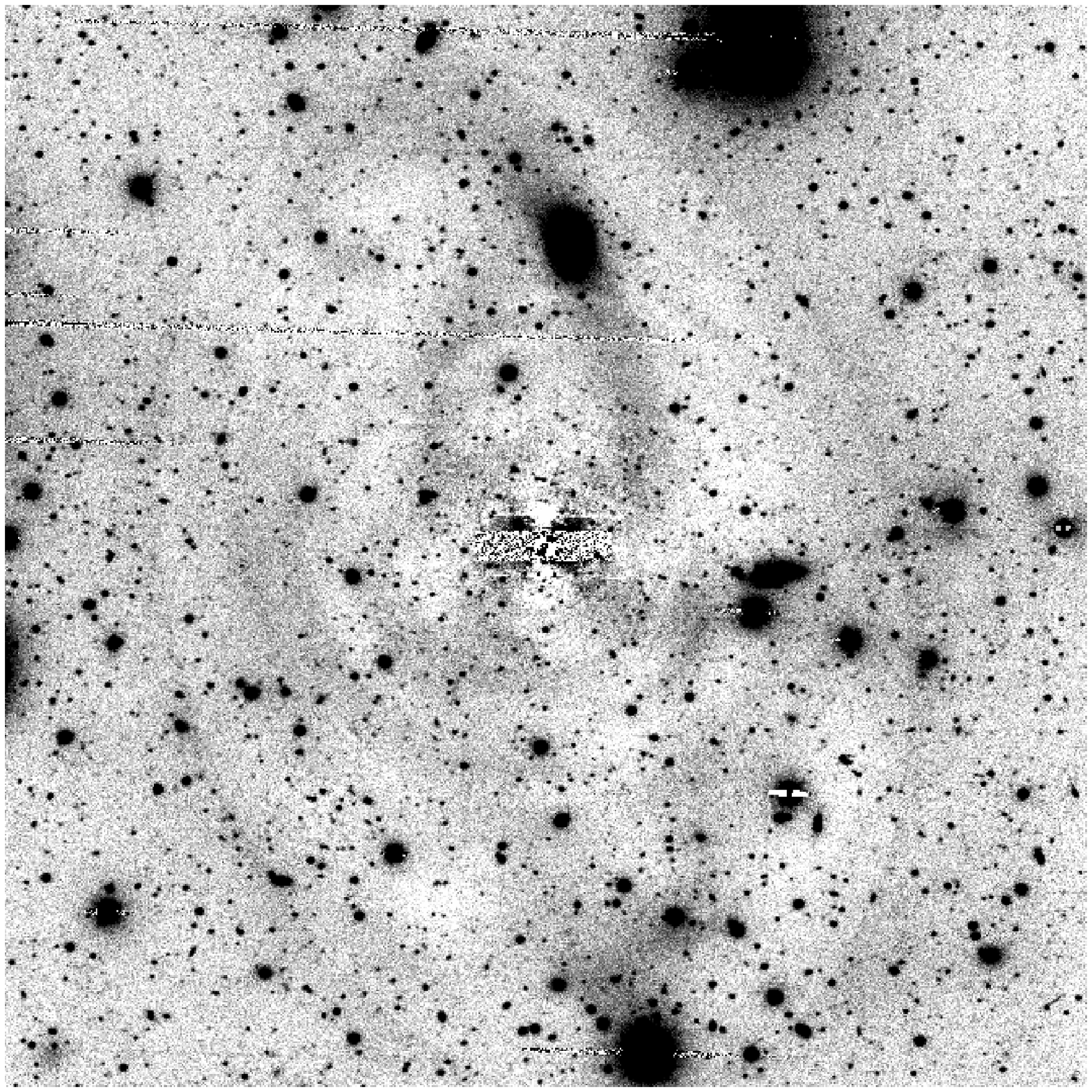}
  \caption{$R$ images of NGC 5812. The upper panel shows the original image
    while the lower panel shows the image after the galaxy light is removed
    with {\it ellipse}, see text. The images are $\sim9'\times9'$. North is up
    and East is to the left.}
  \label{NGC5812galaxyfig}
  \end{centering}
\end{figure}

\subsubsection{Dwarf galaxy interaction with NGC 5812}\label{5812dwarf}

\begin{figure}
  \begin{centering}
    \includegraphics[width=0.48\textwidth]{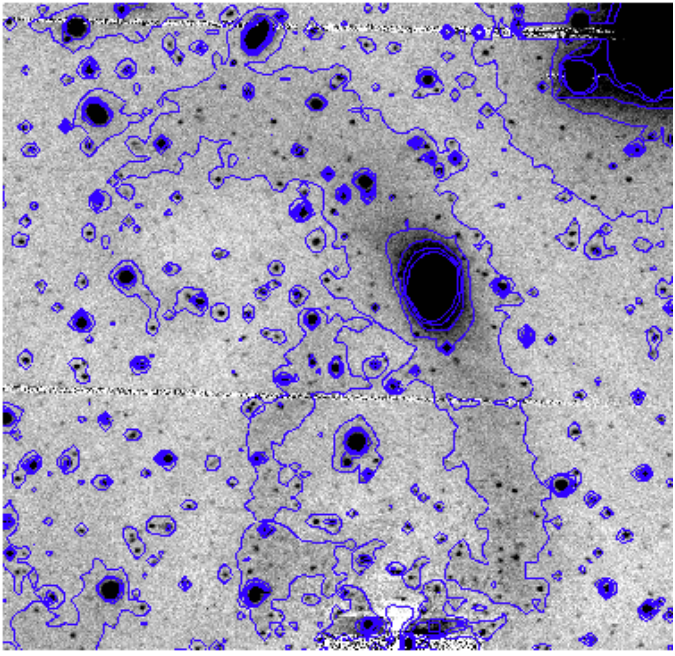}
    \caption{Contour map of the coherent tail structure, in the galaxy
      subtracted $R$-band image, emanating from the interacting dwarf
      galaxy. The lowest contour level (background) is 0.0 increasing in
      steps of 0.1. The image is $\sim4.\!\!'5$ on a side. North is up and East is
      to the left.}
    \label{tailcontours}
  \end{centering}
\end{figure}

It appears that, at least by visual inspection of
Figure~\ref{NGC5812galaxyfig}, that a dwarf companion galaxy is interacting
with NGC 5812; apparent tidal debris can be seen to both the North and South
of the host galaxy in the upper panel and the bottom panel shows what appear
to be tidal arms emanating from the dwarf companion. This dwarf is given the
designation of [MFB03]\,1 by \cite{Madore04} who classify it as a nucleated
(dNE) dwarf. Furthermore, \cite{Tal09} assign NGC 5812 with an interaction
parameter of 0.08, in between their thresholds of marginally and clearly
interacting systems. Although not spectroscopically confirmed,
Figure~\ref{tailcontours} shows a contour map of the structure which clearly
reveals the coherent tidal feature; the tidal nature of the tail is confirmed
by the tell-tail `S' shaped signature of tidal debris and the changing
position angle (PA) of the luminosity contours with radius \cite[e.g.][and
  references therein]{McConnachie06}. It is also clear that the model
subtraction reveals a plethora of other faint structures throughout the
galaxy, an indication that the dwarf must have completed several orbits of NGC
5812.

We wanted to ensure that the observed features were not simply the effects of
flat fielding, or the result of other instrumental or data reduction
processes. To this end, we downloaded the image of NGC 5812 analysed by
\cite{Tal09}\footnote{http://www.astro.yale.edu/obey/index.html}, produced an
     {\it ellipse} model in the same manner as for our own data, and
     subtracted this model from the image. All features seen in our model
     subtracted image in Figure~\ref{NGC5812galaxyfig} are also visible in the
     model subtracted image from the Tal et al. data, including all faint
     structures internal to the galaxy. Therefore, we consider all the
     features revealed in our Figure~\ref{NGC5812galaxyfig} to be real
     structures produced by the tidal stripping of the interacting dwarf.

In addition, an attempt was made to derive the integrated $R$ magnitude of the
dwarf galaxy. This was done with the IRAF {\it ellipse} task, in much the same
way as for the ellipticals themselves. The maximum semi-major axis for the
model was chosen to be $9.5''$ as this was deemed the likely extent of the
main body of the dwarf through visual inspection of the $R$ image (see Table
\ref{apptable3} for the output of the {\it ellipse} model). From the flux
enclosed within the outermost isophote of the model, we calculate the apparent
magnitude for the dwarf galaxy to be m$_R\sim16.6$. Moreover, in addition to
the changing PA of the luminosity contours discussed above, the PA of the
isophotes in the {\it ellipse} model increase steadily with radius from
$\sim-80^\circ$ near the core to $\sim-15^\circ$ in the outer isophotes. This
makes it very clear that the dwarf is undergoing a tidal interaction with NGC
5812, and to the best knowledge of the authors, this is the first definitive
evidence that [MFB03]\,1 is interacting with NGC 5812. See Sections
\ref{discussiondwarf} \& \ref{GCsystems} for further discussion.

\subsection{Point source colour-magnitude diagrams}

\begin{figure}
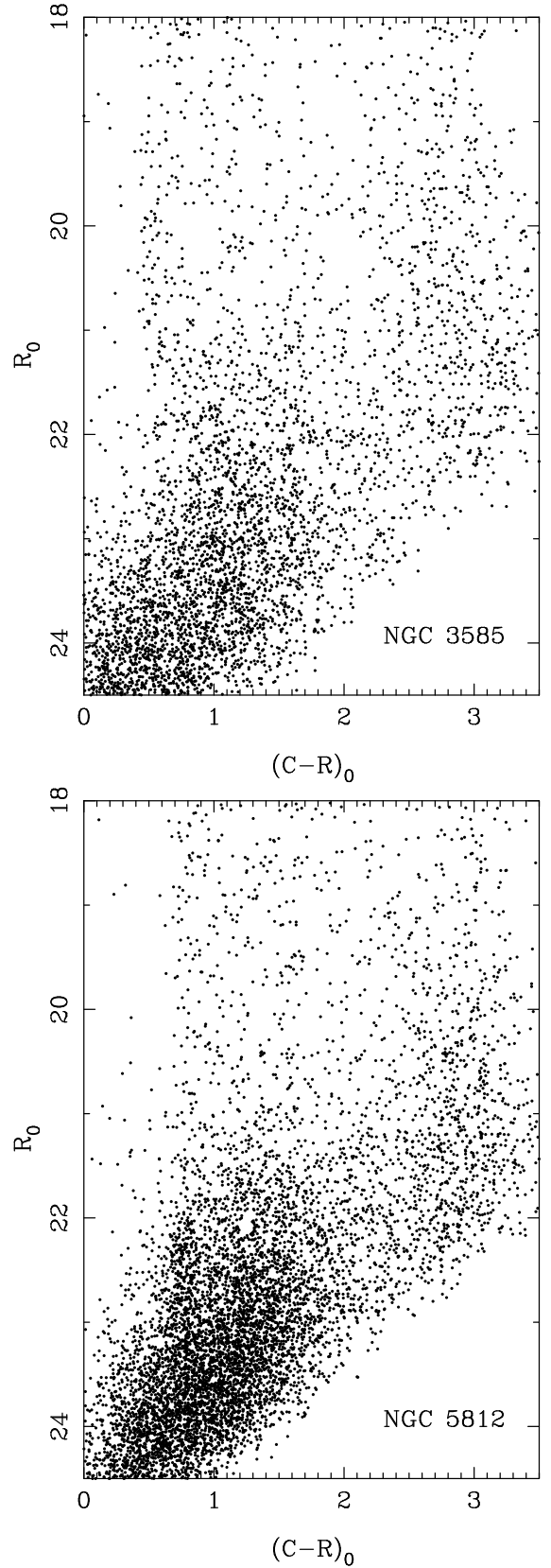

  \begin{centering}
  \includegraphics[angle=-90,width=0.4\textwidth]{figures/NGC3585_CMD_forpaper2.ps}
  \includegraphics[angle=-90,width=0.4\textwidth]{figures/NGC5812_CMD_forpaper.ps}
  \caption{Extinction corrected CMDs of all point sources in the NGC 3585
    (top) and NGC 5812 (bottom) fields. Globular cluster candidates are known
    to have colours in the range $0.9\lesssim (C-R)\lesssim 2.3$
    \citep[e.g.][]{Dirsch03b}. Objects located at $(C-R)\gtrsim 3$ are mostly
    Galactic foreground stars and objects with $R\gtrsim 23$ and
    $(C-R)\lesssim 1$ are mainly background galaxies.}
  \label{CMDfig}
  \end{centering}
\end{figure}

To eliminate the majority of spurious point source detections (those which may
be background galaxies, CCD imperfections, blooming spikes, etc) we performed
cuts based on the $\chi$ and sharpness parameters from the {\it psf} fitting,
restricting our selection to objects with $\chi_C<0.10$, $\chi_R<0.15$,
$-0.8<sharp_C<0.8$ and $-0.8<sharp_R<0.3$. These restrictions were chosen
because it was at these values that the $\chi$ and $sharp$ parameters began to
diverge. This left $\sim5900$ and $\sim8000$ point sources for NGC 3585 and
NGC 5812, respectively, with the brightest, non-saturated, objects having
$R\gtrsim17.5$. These were extinction corrected, taking ${\rm
  E}(B-V)\sim0.064$~and~$\sim0.087$ for NGC 3585 and NGC 5812, respectively
\cite[][]{Schlegel98}. ${\rm E}(B-V)$ values were then converted to ${\rm
  E}(C-T_{\rm 1})$ via the transformation $1.966\times{\rm E}(B-V)={\rm
  E}(C-T_{\rm 1})$ \cite[][]{Geisler91}. Figure~\ref{CMDfig} shows the
colour-magnitude diagrams (CMDs) of all extinction corrected point sources
remaining after the cuts from both fields.

In the CMD from the NGC 5812 field (lower panel of Figure~\ref{CMDfig}) an
apparently extra-Galactic overdensity of point sources can be seen at $R>22$
and $0.6\lesssim(C-R)_0\lesssim0.9$. It is interesting to note that because
this field is in approximately the right location of the sky, the feature at
$R\sim22$ may represent the main sequence turn off (MSTO) of the Sagittarius
(Sgr) dwarf galaxy tidal stream. The Sgr dwarf tails have been shown to have a
main sequence turn off of $V_{\rm MSTO}\sim21$
\cite[e.g.][]{Fahlman96,Majewski99}. Since, in our CMD, these objects are
relatively blue, we expect $(V-R)>0$ and, therefore, if this sequence was the
MSTO of stars in the Sgr dwarf tidal tail we find that $V_{\rm MSTO}>22$. This
is apparently incompatible with the values quoted from the literature and it
must correspond to stars at a greater distance than those in the Sgr tidal
tails. We are, therefore, unsure what this feature in the CMD is likely to be
and leave it as a topic for a future study, as it is outside the scope of this
paper to analyse it in more detail here.

\section{Surface brightness profiles}\label{SBsection}

To determine the surface brightnesses of each galaxy we have chosen to fit a
double beta-model \cite[e.g.][and references therein]{Richtler11} because it
provides an analytic solution for projected luminosity and allows for
analytical deprojection and analytically calculable cumulative luminosity. The
double beta-model is given by:
\begin{equation}
\mu_V(R)=-2.5\times\log\left[a_1\left(1+\left(\frac{R}{r_1}\right)^2\right)^{\alpha_1}+a_2\left(1+\left(\frac{R}{r_2}\right)^2\right)^{\alpha_2}\right].
\end{equation}

\subsection{NGC 3585}

\begin{figure}
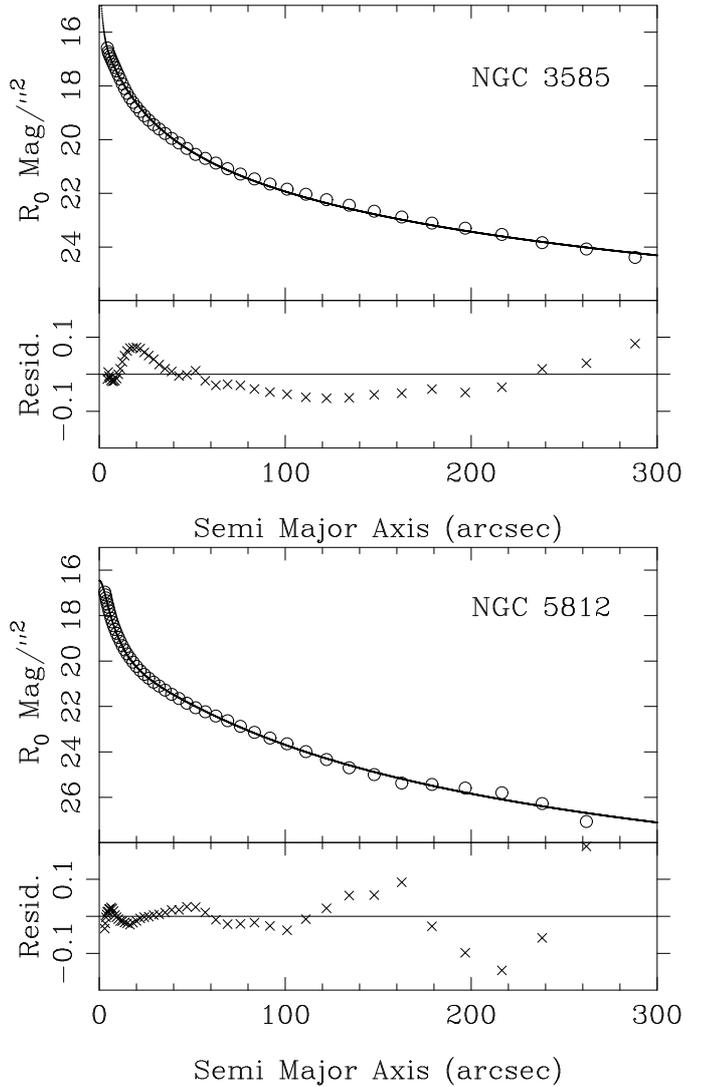

  \begin{centering}
    \includegraphics[angle=-90,width=0.48\textwidth]{figures/NGC3585_SBprofile2.ps}
    \includegraphics[angle=-90,width=0.48\textwidth]{figures/NGC5812_SBprofile2.ps}
    \caption{$R_0$ surface brightness profiles of NGC 3585 (top) and NGC 5812
      (bottom). {\it Top panels:} the circles show the extinction corrected
      $R$ magnitudes per square arcsec, calculated from the $R$ flux as
      measured by the {\it ellipse} task. The solid curves are the double beta
      model fits. The residuals to the fits are shown in the bottom
      panels. Note that in both galaxies the fits are excellent ($\Delta {\rm
        mag}<0.1$) at almost all radii. Data points and residuals are only
      shown for $R>4''$ and $R>3''$ for NGC 3585 and NGC 5812, respectively,
      due to the limitations of the {\it ellipse} models in the central
      regions.}
    \label{3585SBfig}
  \end{centering}
\end{figure}

Our best fit double beta-model to the flux of NGC 3585 (from the {\it ellipse}
model) has $a_1=2.0\times10^{-6}$, $r_1=1.0''$, $a_2=1.5\times10^{-7}$,
$r_2=10.0''$ and $\alpha_1=\alpha_2=-1.0$. The double-beta surface brightness
model and the output of the {\it ellipse} task, both in magnitudes per square
arcsecond, are shown together in Figure~\ref{3585SBfig}.

\subsection{NGC 5812}

Treating NGC 5812 in the same manner as NGC 3585 we find that the best fitting
double beta-model has $a_1=2.6\times10^{-7}$, $r_1=3.9''$,
$a_2=3.3\times10^{-9}$, $r_2=55.2''$ and $\alpha_1=-1.2$ and $\alpha_2=-1.9$
(Figure~\ref{3585SBfig}).

Note that our surface brightness models in Figure~\ref{3585SBfig} are fits to
the semi-major axis of the isophotes. Therefore, to test of the veracity of
our surface brightness models, we also produced fits to the flux enclosed
within circular apertures and integrated these out to various radii to
determine the enclosed flux. We then converted these into magnitudes for
comparison with \cite{Prugniel98}. We find that at all available radii, our
surface brightness models for both galaxies agree to within $0.2$ mag to the
values quoted by Prugniel, however, the agreement is almost always better than
$0.05$ mag. For those radii where our integrated surface brightness model
magnitudes differ from those by Prugniel the most, the Prugniel values are
apparently spurious (e.g. for NGC 3585, the quoted magnitude for a $\sim13''$
radius aperture is brighter than for apertures of larger radii).

\section{$(C-R)$ colour maps \& colour profiles}\label{coloursection}

Colour maps of elliptical galaxies can show, for example, evidence of dust and
star formation, by revealing minute differences in colour in the luminous
component of a galaxy \cite[e.g. in NGC 1316][]{Richtler12}. We produced
colour maps for both galaxies, with special interest in searching for any
interesting colour variations in the central discy feature of NGC
3585.

\begin{figure}
  \begin{centering}
    \includegraphics[width=0.48\textwidth]{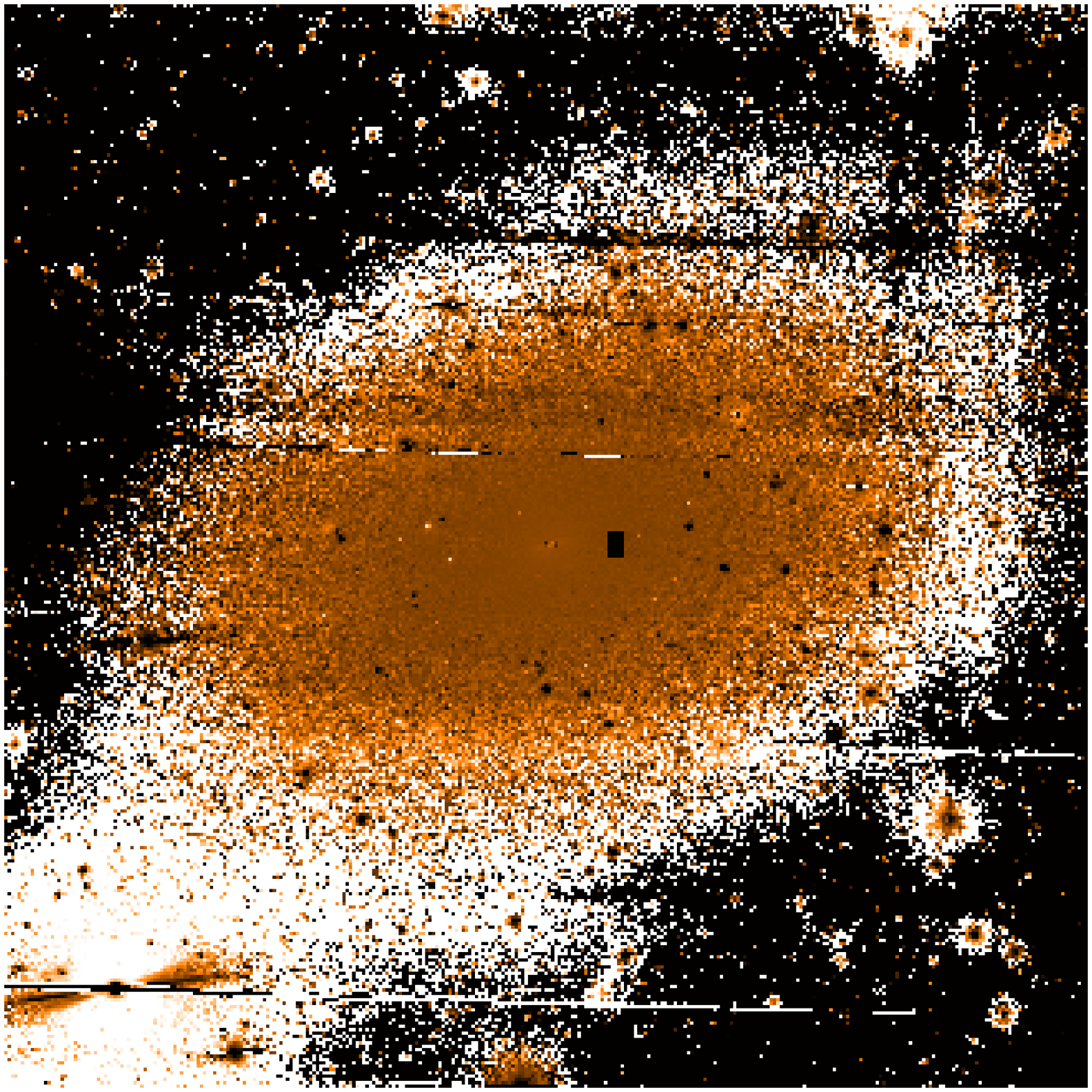}
    \includegraphics[width=0.48\textwidth]{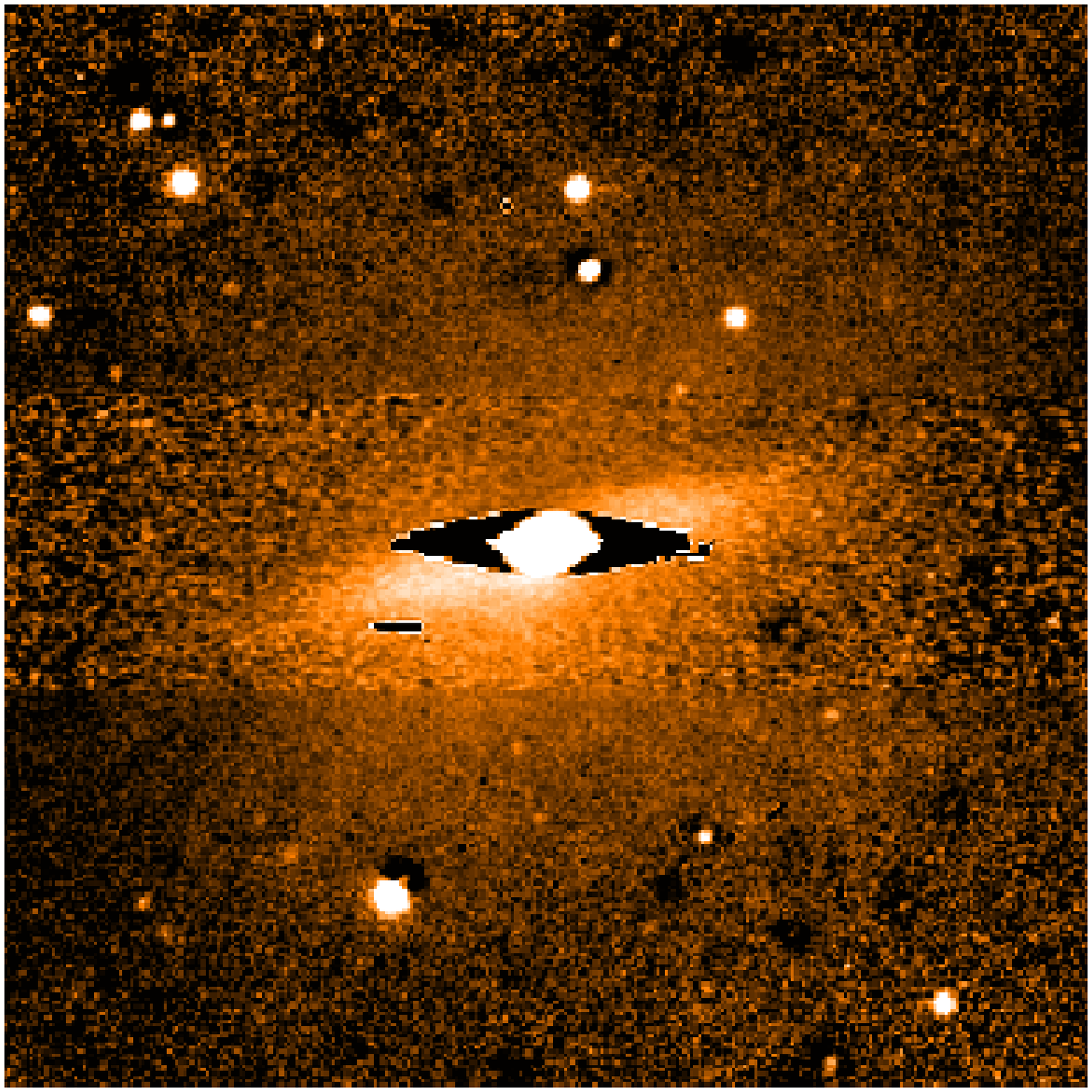}
    \caption{The $(C-R)$ colour maps of NGC 3585. The top panel shows the
      entire galaxy and has the same dimensions as Figure
      \ref{NGC3585galaxyfig} ($\sim15'\times15'$). The colour range shown is
      $1.1\lesssim(C-R)\lesssim2.5$, with bluer colours darker, and redder
      colours lighter. It is clear that no strong colour gradient is present
      out to large radii. The lower panel shows the inner $\sim75''\times75''$
      over a colour range of $1.73\lesssim(C-R)\lesssim1.85$. The $\sim45''$
      discy structure discussed in the text is clearly visible. The noisy
      rectangular region surrounding the galaxy core is the short exposure
      that was used to replaced the overexposed central region (see Section
      \ref{galmodsec}) and does not affect the colour profile in any way.}
    \label{NGC3585colourmaps}
  \end{centering}
\end{figure}

\begin{figure}
  \begin{centering}
    \includegraphics[width=0.48\textwidth]{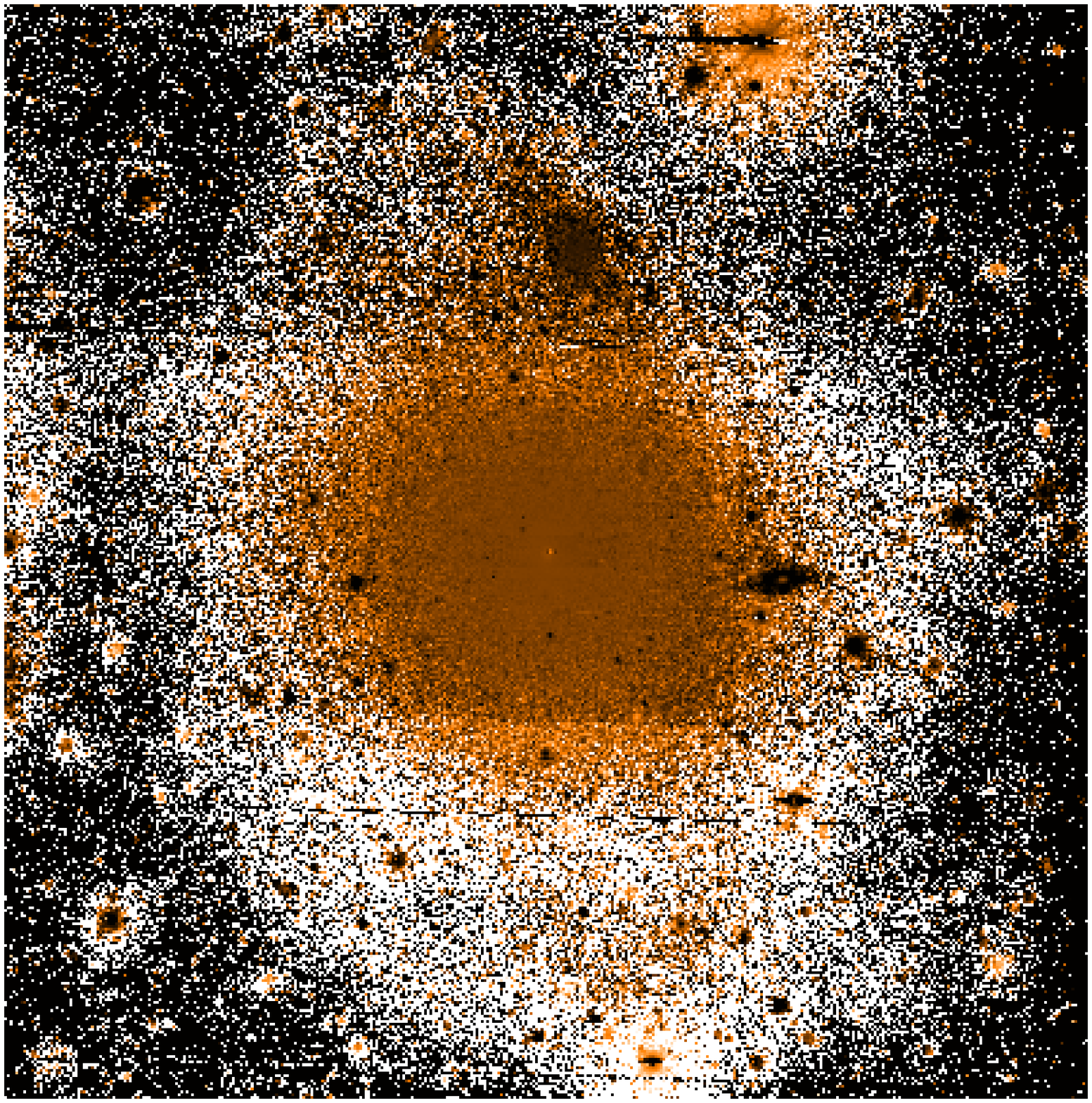}
    \includegraphics[width=0.48\textwidth]{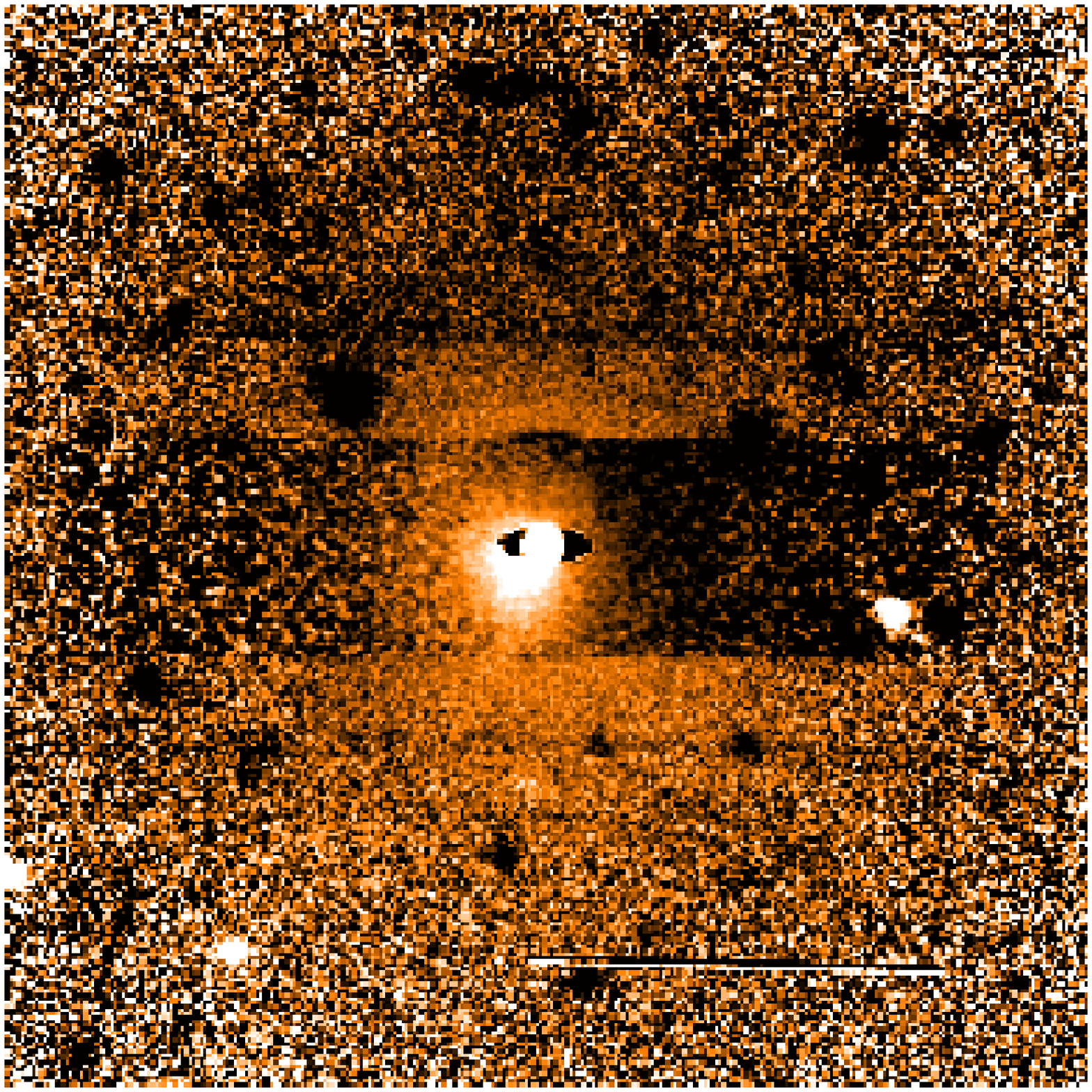}
    \caption{The $(C-R)$ colour maps of NGC 5812. The top panel shows the
      entire galaxy and has the same dimensions as Figure
      \ref{NGC5812galaxyfig} ($\sim9'\times9'$). The colour range shown is
      $1.1\lesssim(C-R)\lesssim2.5$, with bluer colours darker, and redder
      colours lighter. It is clear that no strong colour gradient is present
      out to large radii. The lower panel shows the inner $\sim75''\times75''$
      over a colour range of $1.71\lesssim(C-R)\lesssim1.80$. The circular,
      redder region, $\sim15''$ in diameter, as discussed in the text, is
      clearly visible. The noisy rectangular region surrounding the galaxy
      core is the short exposure that was used to replaced the overexposed
      central region (see Section \ref{galmodsec}) and does not affect the
      colour profile in any way.}
    \label{NGC5812colourmaps}
  \end{centering}
\end{figure}

\begin{figure}
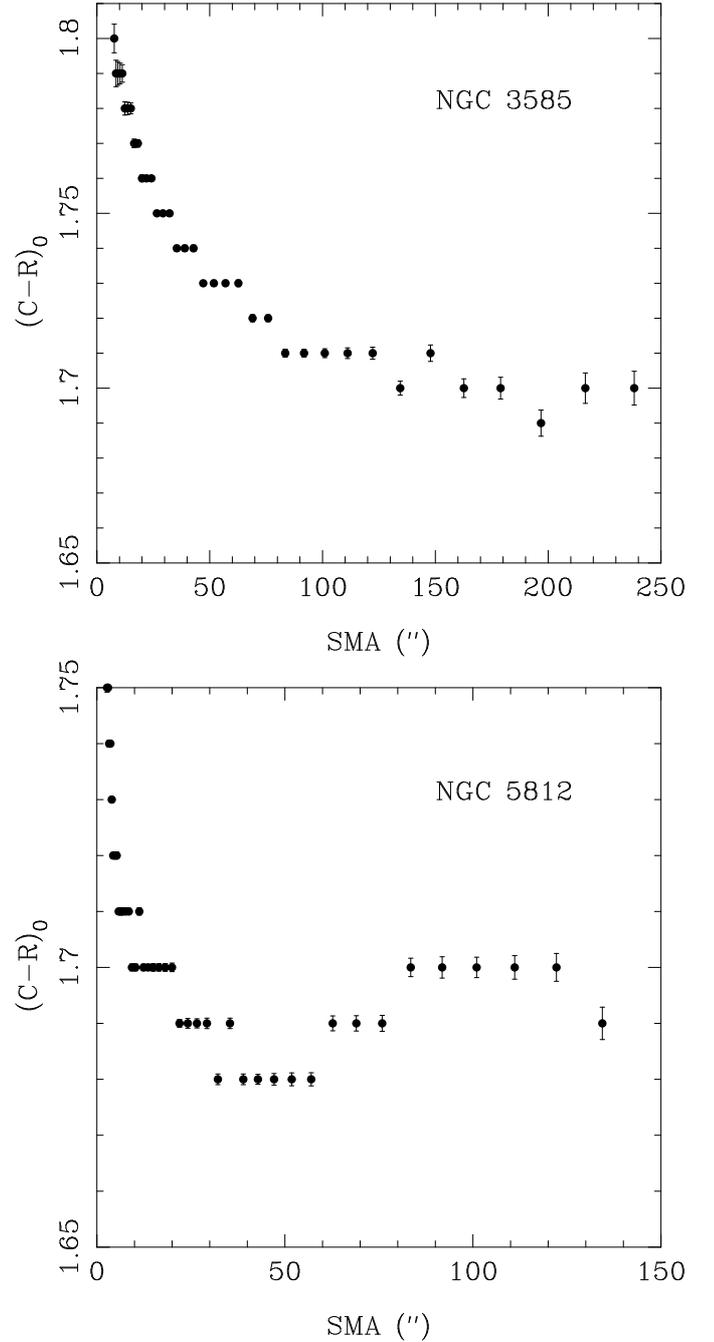

  \begin{centering}
    \includegraphics[angle=-90,width=0.48\textwidth]{figures/NGC3585_colourprofile_final.ps}
    \includegraphics[angle=-90,width=0.48\textwidth]{figures/NGC5812_colourprofile_final.ps}
    \caption{Colour profiles of NGC 3585 (top) and NGC 5812 (bottom). Note the
      trend to redder colours at small radii in both galaxies. The inner
      region of NGC 3585 is dominated by discy structure (see
      Figure~\ref{NGC3585galaxyfig}) which has a slightly redder colour than
      the mean. The very inner region of NGC 5812 contains a slightly
      off-centre, almost spherical, redder region. Note that in both cases the
      colour of the reddest, inner-most regions is $\lesssim0.1$ mag redder
      than the mean colour of the galaxy.}
    \label{colourprofiles}
  \end{centering}
\end{figure}

The colour maps were produced by converting the reduced, combined, extinction
corrected images to the magnitude scale with the IRAF routine {\it imarith}
(i.e. $-2.5\times {\rm log[flux]}$) for both filters. Note that the THELI
pipeline does not estimate sky/background uncertainties so we estimated these
uncertainties using Sextractor \cite[][]{Bertin96}. We used a BACK\_SIZE of 64
and BACK\_FILTERSIZE of 7 to ensure that all faint objects were extracted. The
background estimates were very nearly identical to those given by the THELI
pipeline, within $\sim2$ counts, a good indication that our background
estimates are reliable. The background uncertainties from Sextractor are
$\sim11.5$ and $\sim15.9$ counts, respectively, for the $C$ and $R$ images of
NGC 3585, and $\sim12.1$ and $\sim15.5$ counts, respectively, for the $C$ and
$R$ images of NGC 5812, which is the $1\sigma$ deviation from the mean of all
pixels considered background. We then subtracted the $R$ image from the $C$
image to produce a 2D $(C-R)$ colour map (Figures \ref{NGC3585colourmaps} \&
\ref{NGC5812colourmaps}). The {\it ellipse} task was then employed to produce
``colour isophotes'' (or isochromes) in the same manner as for the galaxy
surface brightness models. The isochromes were then converted to colour
profiles (Figure~\ref{colourprofiles}).

The most interesting result from the 2D colour maps is that we found almost no
evidence for any dust, or star formation, within either galaxy. NGC 3585 shows
a small, edge on, slightly redder disc $\sim45''$ across in the centre of the
galaxy. This coincides perfectly with the position angle of the discy core
discussed in Section~\ref{galmodsec}. The most obvious explanation for this
reddening is dust, however, the reddening is very slight ($\sim0.1$ mag, see
Figures~\ref{NGC3585colourmaps} \& \ref{colourprofiles}), therefore, if dust
is the cause, it must be very diffuse.

NGC 5812, on the other hand, has a small ($\sim15''$ diameter) circular
reddened region in its core, however, the reddening in this case is much lower
($\sim0.05$ mag) so, again, the dust, if this is the cause, must be even more
diffuse than that in the core of NGC 3585. The dwarf companion of NGC 5812 has
a mean colour of $(C-R)_0\sim1.3$, i.e. $\sim0.4$ bluer than NGC 5812,
however, there is no obvious blue excess associated with tidal tails of the
dwarf in either the colour maps, or in Figure~\ref{colourprofiles}. It is our
opinion that this lack of obvious colour excess in the tidal tails is due to
our photometry not being deep enough or that the bright body of NGC 5812
simply overwhelms the faint tidal tails to such a degree that the colour of
the tails are lost.

Outside the inner regions, the colour profiles of both galaxies are remarkably
flat, both having an intrinsic colour of $(C-R)\sim1.7$, a colour fairly
typical for elliptical galaxies. Interestingly, while the colour of NGC 3585
is fairly typical for elliptical galaxies, it is unclear how the youth of the
galaxy and the blue component of the GC population can be reconciled with the
galaxy colour.

\section{The globular cluster systems}\label{GCsection}

\begin{figure}
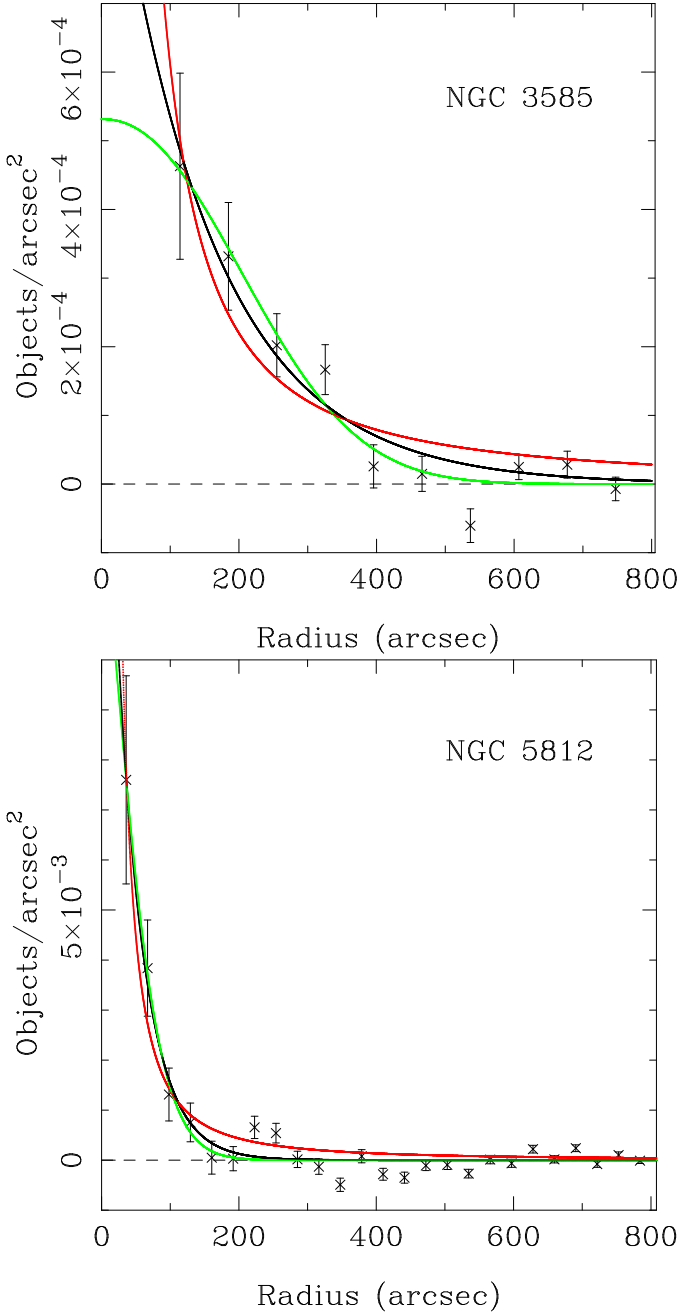

  \begin{centering}
  \includegraphics[angle=-90,width=0.48\textwidth]{figures/NGC3585_density_subtracted_10bins+fits.ps}
  \includegraphics[angle=-90,width=0.48\textwidth]{figures/NGC5812_density+fits.ps}
  \caption{Background subtracted surface density of point sources per square
    arcsecond within $800''$ of NGC 3585 (top) and NGC 5812 (bottom). The
    density falls to the background level at $R\sim400''$ for NGC 3585 and
    $R\sim200''$ for NGC 5812. The black, green and red curves are the best
    fit exponential, S\'ersic and power law functions, respectively. In both
    cases the S\'ersic drop offs describe the density profiles much more
    closely than either a power law or an exponential.}
  \label{GCsurfacedensity}
  \end{centering}
\end{figure}

To determine the point source background densities we binned all point sources
within each field into annular bins to produce a density profile. The
background was then located visually and subtracted. The results are shown in
Figure~\ref{GCsurfacedensity}. The surface density of the NGC 3585 field drops
to the background ($3.9\times10^{-4}$ point sources per square arcsecond) at
$R_{\rm bg}\sim400''$ and for NGC 5812 at $R_{\rm bg}\sim200''$
($4.8\times10^{-3}$ point sources per square arcsecond).  Furthermore, it is
clear that the distributions of GC candidates of both galaxies do not follow a
strict power law and are, in fact, best described by a S\'ersic profile
(Figure~\ref{GCsurfacedensity}).

\begin{figure}
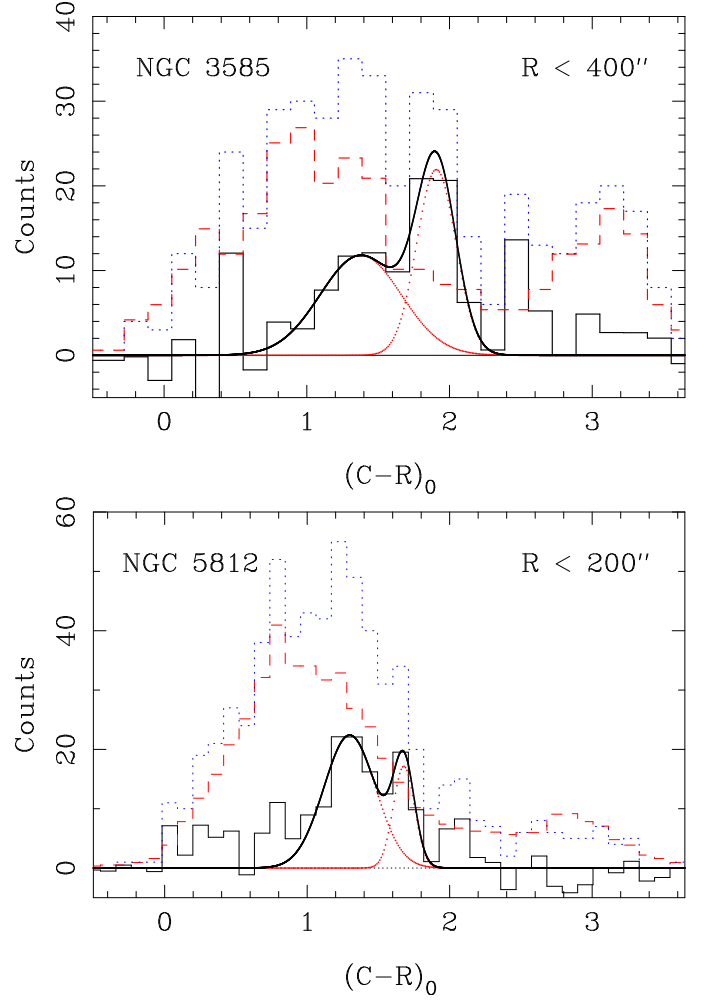

  \begin{centering}
  \includegraphics[angle=-90,width=0.48\textwidth]{figures/NGC3585_hist_lt-gt_norm_36bins+0.263_forpaper.ps}
  \includegraphics[angle=-90,width=0.48\textwidth]{figures/NGC5812_hist_lt-gt_norm+0.17_forpaper.ps}
  \caption{Area normalised, extinction corrected histograms of point sources
    from the NGC 3585 (top) and NGC 5812 (bottom) fields. The blue dotted
    histograms are all sources with radii $<400''$ for NGC 3585 and radii
    $<200''$ for NGC 5812, i.e. all sources at radii with point source counts
    higher than the background (see Figure~\ref{GCsurfacedensity}). The red
    dashed histograms show normalised counts for all sources with radii
    $>540''$ for NGC 3585 and radii $>400''$ for NGC 5812, ensuring that point
    source counts have dropped to the background level. The solid black
    histogram shows the inner counts minus the background. Double Gaussian
    fits to the two main peaks are shown as solid black curves, and the red
    dotted curves are the individual Gaussian components. The double Gaussian
    peaks are at $(C-R)\sim1.37$ ($\sigma\sim0.28$) and $(C-R)\sim1.91$
    ($\sigma\sim0.14$) for NGC 3585 and $(C-R)\sim1.30$ ($\sigma\sim0.18$) and
    $(C-R)\sim1.68$ ($\sigma\sim0.07$) for NGC 5812. Bin widths are 0.17\,mag
    for NGC 3585 and 0.11\,mag for NGC 5812.}
  \label{GChist}
  \end{centering}
\end{figure}

To properly characterise the GC candidates it is necessary to produce colour
distributions of the point sources within the radii determined above.
Histograms of the normalised background sources, sources with $R<R_{\rm bg}$
and the difference between the two are shown in Figure~\ref{GChist}. Old GC
populations appear to have a Universal colour peak at $(C-T_1)\sim1.3$
\cite[e.g.][]{Dirsch03b,Dirsch05,Lee08,Park10,Schuberth10,Richtler12}.
Figure~\ref{GChist} shows that both NGC 3585 and NGC 5812 also contain
populations with this colour peak, strengthening this apparent
universality. The strength of the blue peak in NGC 5812, in comparison the
weaker blue peak of NGC 3585, may be explained by accretion of old GCs from
the interacting dwarf galaxy. The blue GCs are not obviously concentrated near
the dwarf, however, this is not surprising because the dwarf has obviously
been through several orbits, based on the length of the tidal tails, and the
plethora of faint tidal debris throughout the elliptical
(Section~\ref{5812galmod}). It is, therefore, to be expected that these
accreted GCs should, at the present time, be fairly evenly distributed
throughout the galaxy.

Both GC populations exhibit blue excesses -- objects with bluer colours than
those of the old GCs ($[C-R]\lesssim0.8$). These may be young GCs. In the case
of NGC 3585, the $\sim12$ objects with $(C-R)\sim0.5$ may be a young
population, of unknown origin. In the case of NGC 5812 these objects have a
large spread in colour $0\lesssim(C-R)\lesssim0.8$, evidence for an extended
formation period. We discuss the possibilities for the blue GC populations of
both galaxies in Section~\ref{conclusions}. Both galaxies also exhibit some
red excess, with objects redder than that of the main red peaks,
i.e. $(C-R)\gtrsim2.4$ and $(C-R)\gtrsim1.9$ for NGC 3585 and NGC 5812,
respectively. For both galaxies these red objects are scattered fairly evenly
throughout the field, so cannot be attributed to a background galaxy
cluster. However, for NGC 5812, there are only a few objects, and these may be
individual background galaxies. For NGC 3585, the vast majority of the red
excess is concentrated in a small colour range,
$2.4\lesssim(C-R)\lesssim2.8$. The nature of these objects is unknown and,
therefore, for the time being, remains a mystery.

It is interesting to note that the ratios of the full width half maxima (FWHM)
of the Gaussian fits to the GC candidate blue and red peaks in each of the
fields is nearly identical at $\sim2.0$ despite the other parameters differing
significantly. This is not a Universal property of elliptical galaxy GC
populations, and this ratio may be useful in determining their evolutionary
histories. As the GC populations of more elliptical galaxies are studied it
should become clear what can be gleaned about their evolutionary histories
from this information.

\subsection{NGC 3585}\label{3585GCs}

\begin{figure}
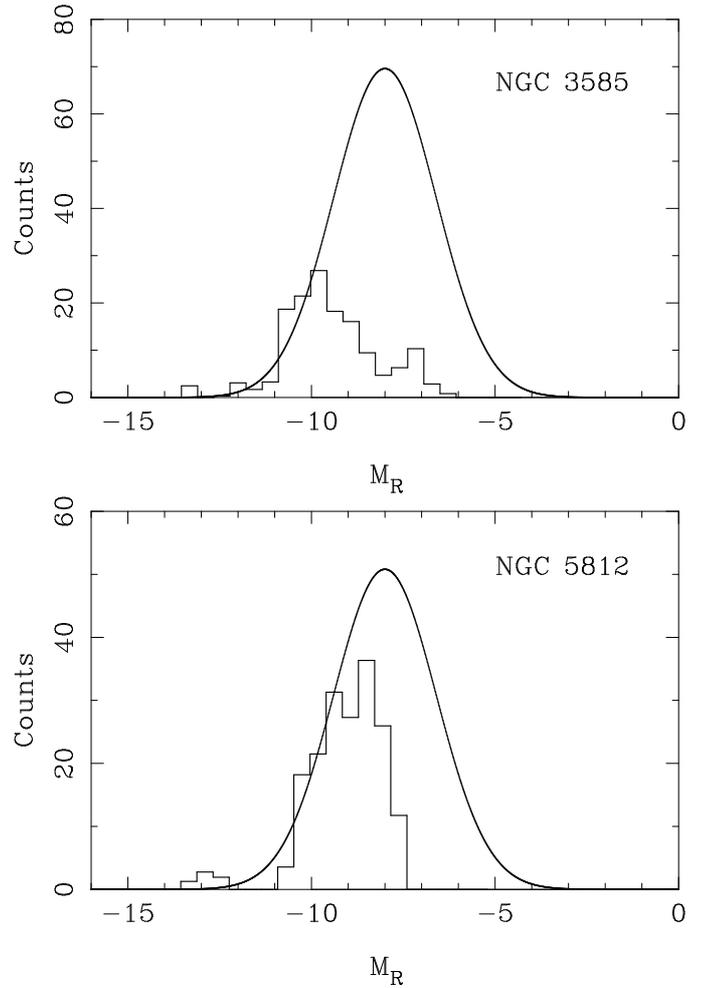

  \begin{centering}
  \includegraphics[angle=-90,width=0.48\textwidth]{figures/NGC3585_R_luminosityftn.ps}
  \includegraphics[angle=-90,width=0.48\textwidth]{figures/NGC5812_R_luminosityftn.ps}
  \caption{Histograms of the background subtracted luminosity functions of the
    GC candidates of NGC 3585 (top) and NGC 5812 (bottom) in absolute $R$
    magnitudes. The solid curves are Gaussian fits to the data, restricting
    the mean to M$_R=-8.0$ and the width to $\sigma=1.4$, and fitting only to
    complete magnitudes, as described in the text. Bin widths in both cases
    are 0.44\,mag.}
  \label{lmftn}
  \end{centering}
\end{figure}

In total $\sim130$ GC candidates have been identified in the NGC 3585
field. The luminosity function (GCLF) of the GC candidates is shown in
Figure~\ref{lmftn}. The GCLF was derived by binning the $R$ magnitudes of
point sources with $0.6\lesssim(C-R)\lesssim2.4$, of both background sources
(those at radii larger than $540''$), and sources at radii less than $400''$
and subtracting one from the other, in a similar manner as the colour
profile. A Gaussian fit was produced by fixing the $R$ turnover magnitude at
M$_{R,TO}=-8.0$ (assuming M$_{V,TO}\sim-7.4$ [\citealt{Rejkuba12}] and
$[V-R]\sim0.6$ [Table~\ref{paramtable}]). Because we have not performed a
completeness test on our point source data, we fixed the width of the Gaussian
fit to be $\sigma=1.4$ \cite[][and references therein]{Rejkuba12} to ensure
that the uncertainty in the total number of GCs was minimised as much as
possible. The amplitude was left as the only free parameter. The best fit has
an amplitude of $\sim69.6$. Integrating this Gaussian gives the estimated
total number of GCs expected in the system to be $\sim550$. Note that this
Gaussian was only fit to objects brighter than $R=21$ where our data is close
to complete. It should be noted here that we did not perform any completeness
testing because the only result relying on the completeness of the data is the
total number of GCs based on the GCLF. We note that this lack of a
completeness test will affect the uncertainty of the calculated number of GCs
and that this is, therefore, only and indication of the true number. We note
also, however, that from previous experience
\cite[e.g.][]{Conn07,Conn08,Conn12} we can be fairly certain that the
completeness is close to 100\% at the limiting magnitude we have chosen
($R=21$).

Assuming the specific frequency relation given by \cite{Harris91} and an
absolute magnitude of M$_V=-21.8$ from Section~\ref{galmodsec} gives a
specific frequency of $S_N\sim1.05$. This is, it must be stressed, only an
indication of the true specific frequency since our total number of GCs is
only an estimate. However, this is fairly typical, albeit at the low end, for
specific frequencies of elliptical galaxies in low density environments
\cite[e.g.][]{Harris91}. It should be noted that the variation in distance
estimates quoted by NED (see Table \ref{paramtable}) will affect specific
frequency estimate. Using the lower and upper limits for the distance quoted
by NED gives absolute magnitudes of M$_V=-21.1$ and M$_V=-22.4$,
respectively. This leads to specific frequencies of $S_N\sim2.0$ and
$S_N\sim0.6$, however, we remind the reader that our estimates are only an
indication of the true specific frequency due to the uncertainties in our
estimates for the total number GCs, as well as the distance uncertainties in
the literature. It should also be noted that this range of specific
frequencies is still within the accepted range for IEs
\cite[e.g.][]{Harris91}. Furthermore, the most recent distance estimates
mostly cluster around our adopted distance of $\sim18.3$\,Mpc. This indicates
that the large discrepancy in distances quoted by NED is probably misleading,
and that both our adopted distance for this galaxy, and hence the specific
frequency, are more reliable that it may appear.

\cite{Humphrey09}, using Wide Field Planetary Camera 2 on the {\it Hubble
  Space Telescope} ({\it HST}), observed 102 GCs candidates associated with
NGC 3585 and found M$_{V,TO}\sim-6.75$ and $S_N\sim0.47$. This $\sim1$ mag
difference between the turnover measured by \cite{Humphrey09} and that quoted
for old GCs by \cite{Rejkuba12}, and the $\sim0.5$ discrepancy between the
specific frequency quoted by \cite{Humphrey09} and our measurement, can
possibly be attributed to the small radii analysed by \cite{Humphrey09} and,
of course, to the uncertainty in our estimate of $S_N$. The largest radii
measured by \cite{Humphrey09} is $1.\!\!'5$, which is $<20$\% of the radius at
which we find GC candidates. Moreover, when including all $\sim1000$ GCs in
their sample (from 19 early-type galaxies) \cite{Humphrey09} find
M$_{V,TO}\sim-8$ (note that the distance adopted by Humphrey is
$18.6\pm\sim2$\,Mpc and ours is 18.3\,Mpc, so our adopted distance is well
within the uncertainties of the Humphrey value, therefore, M$_{V,TO}\sim-8$ is
a reasonable assumption). However, it is very possible that the GC population
of NGC 3585 is young, which would account for the dimmer peak in the GCLF (see
Section~\ref{conclusions}). A young GC population does not, however, account
for the low $S_N$ value stated by \cite{Humphrey09}. Whatever the cause of the
discrepancy between our $S_N$ and that quoted by \cite{Humphrey09}, the peak
absolute magnitude of our GCLF does not affect the shape of our Gaussian fit
to the GCLF nor, therefore, our $S_N$ or estimated total number of GCs.

Two studies list catalogues of GC candidates for the NGC 3585 system, namely
\cite{Hempel07}, with 26 candidates using the FORS2 instrument on the Very
Large Telescope, and \cite{Humphrey09} with 102 using {\it Hubble Space
  Telescope} data from the WFPC2 instrument. Cross matching our GC candidates
with those catalogues finds only two matches. This paucity in common GC
candidates is due those studies being restricted to the inner $\sim1.\!\!'5$
(see Figure \ref{NGC3585galaxyfig} for the approximate size of the footprint
of WFPC2, as well as the approximate radius beyond which our point source
counts drop to the background level, overlaid onto our $R$ image) as well as
to faint objects. The two GC candidates in common with those studies are shown
in Table~\ref{GCcommontable} which lists positions and magnitudes in available
filters from the literature and the current work.

\begin{table*}[!ht]
\centering
\begin{threeparttable}
\caption{Globular cluster candidates of NGC 3585 in common with literature
  values}\label{GCcommontable}
\begin{tabular}{@{}cccccccc@{}}
\hline
\hline
ID & RA & dec & $V$ (mag) & $I$ (mag) & $K$ (mag) & $C$ (mag)$^{(c)}$ & $R$ (mag)$^{(c)}$\\
\hline
$6^{(a)}$ & $11^{\rm h}13^{\rm m}13.8^{{\rm s} (a,c)}$ &
$-26^\circ45'51.1''^{(a,c)}$ & $22.70\pm0.02^{(a)}$ & $20.55\pm0.01^{(a)}$
& $18.60\pm0.04^{(a)}$ & $22.75\pm0.05$ & $20.58\pm0.01$\\
$85^{(b)}$ & $11^{\rm h}13^{\rm m}19.1^{{\rm s} (b,c)}$ &
$-26^\circ44'16.9''^{(b,c)}$ & $22.35\pm0.03^{(b)}$ & $21.31\pm0.02^{(b)}$ &
--  & $23.35\pm0.06$ & $21.10\pm0.02$\\
\hline
\end{tabular}
\begin{tablenotes}
\item[{\it a}] \cite{Hempel07}
\item[{\it b}] \cite{Humphrey09}
\item[{\it c}] {This work}
\end{tablenotes}
\end{threeparttable}
\end{table*}

\subsection{NGC 5812}

In total $\sim180$ GC candidates have been identified in the NGC 5812
field. As with NGC 3585, the luminosity function of the GC candidates is shown
in Figure~\ref{lmftn}. The luminosity function was derived in the same manner
as NGC 3585, however, the colour range was chosen to be
$0.8\lesssim(C-R)\lesssim2.0$, the inner radius to be $200''$ and the radius
at which point sources drop to the background level was $400''$. Again the
turnover magnitude for the Gaussian fit was chosen to be M$_R=-8.0$ and the
width was chosen to be $\sigma=1.4$, with the amplitude left free. The best
fit has an amplitude of $\sim50.8$. Integrating this Gaussian in the same
manner as for NGC 3585 gives the estimated total number of GCs expected in the
system as $\sim400$. Note that this Gaussian was only fit to objects brighter
than $R=21$ to ensure we were fitting to complete data.

Again, assuming the specific frequency relation by \cite{Harris91}, we find a
specific frequency of $\sim1.2$. This is, as with NGC 3585, only an indication
of the true specific frequency, however, it is fairly typical for specific
frequencies of elliptical galaxies in low density environments
\cite[e.g.][]{Harris91}. As for NGC 3585, if we assume the extremes of
distance estimates from NED, we find absolute magnitudes of M$_V\sim-21.0$ and
M$_V\sim-21.7$ giving $S_N\sim1.6$ and $S_N\sim0.8$, respectively. Again, as
with NGC 3585, these values are well within the accepted range for IEs, with
the uncertainties arising from the uncertainties in both literature distances
and our estimate of the total number of GCs. Note also that, as with NGC 3585,
the more recent distance estimates all cluster around the mean value we have
adopted ($\sim27.7$\,Mpc) indicating that both our adopted distance for this
galaxy, and hence the specific frequency, are more reliable that it may appear
at first glance.

\section{Discussion and Conclusions}\label{conclusions}

Using wide-field Washington $C$ and $R$ photometry, reaching to a point source
magnitude of $R\sim24$, we have produced accurate galaxy models for NGC 3585
and NGC 5812, using the IRAF {\it ellipse} task. In this Section we discuss
the analysis of these models, and the underlying structures revealed through
their subtraction. Note that all parameters derived in the current work are
presented in Table~\ref{derivedparams} for clarity.

\begin{table*}[!ht]
\centering
\begin{threeparttable}
\caption{Parameters for each galaxy derived in the current paper}\label{derivedparams}
\begin{tabular}{@{}ccccccccc@{}}
\hline
\hline
Name & m$_R$ & $R_e$ & $\Sigma R_e$ & $<(C-R)>$ & $N_{\rm GC}$ & $S\!_N$ & \\
 & (mag) & ($''$) & ($R$ mag$''^{-2}$) & (mag) & \\
\hline
NGC 3585 & 8.9 ($<380''$) & $80$ & 6.7 & 1.70 & 550 & 1.05 & \\
NGC 5812 & 10.1 ($<270''$) & $35$ & 8.0 & 1.69 & 400 & 1.20 & \\
\hline
\end{tabular}
\begin{tablenotes}
\item The columns are: Galaxy name, apparent $R$ magnitude (within the quoted
  semi-major axis), effective radius in arcseconds, $R$ surface brightness
  within $R_e$ in magnitudes per square arcsecond, mean $(C-R)$ colour of the
  galaxy light, estimated total number of GCs (based on the GCLF) and specific
  frequency.
\end{tablenotes}
\end{threeparttable}
\end{table*}

\subsection{Galaxy Morphologies}

We find that the outer isophotes of NGC 3585 become slightly less elliptical
at large radii, which is expected for a galaxy with a discy core. After
subtracting the galaxy model, an unexplained $\cap$-shaped feature surrounding
the centre of NGC 3585 is revealed. This feature has the same colour (to
within 0.01 mag) as the surrounding galaxy light, and we do not know the
nature of this feature. It is beyond the scope of this paper to investigate
this further, however, we are under the working assumption that is a real
feature since we have no reason to doubt the veracity of our galaxy
model. After subtracting the models, many other structures are also revealed
in both the NGC 3585 and NGC 5812 systems, indicative of complex dynamical
systems.

The surface brightness profiles for each galaxy were fitted with double
beta-models, the integration of which result in magnitudes that agree very
well with literature values. Two dimensional colour maps, and profiles, were
produced and show a distinct lack of any star formation, or dust (except in
the very centres of the galaxies), which seems to indicate no recent major
merger activity. In addition, the very slight blue colour trend for the minor
axis and red, $\sim40''$ diameter, disc in the core of NGC 3585, originally
shown by \cite{Fisher96}, are both confirmed. Our colour profiles, and 2D
colour maps, reveal a red discy region with a semi-major axis of $\sim45''$,
probably associated with diffuse dust, in the core of NGC 3585. A fairly
spherical region of very diffuse dust also surrounds the core of NGC 5812, out
to a radius of $\sim7-8''$. Outside the diffuse dust in the central regions,
the colour profile of each galaxy is very flat, and is fairly typical of field
ellipticals at $(C-R)\sim1.7$.

\subsection{Dwarf galaxy interaction with NGC 5812}\label{discussiondwarf}

The interaction between NGC 5812 and its dwarf companion galaxy, [MFB03]\,1,
has been confirmed here for the first time. The coherent nature of the debris
resulting from the interaction, as well the change in position angle of the
isophotes with radius, and the tell-tail `S' shape of the isophotal contours
of the region immediately surrounding the dwarf, confirm the tidal nature of
this debris \cite[][and references therein]{McConnachie06}. We have calculated
the apparent magnitude of the dwarf to be m$_R\sim16.6$.

While it is clear that the dwarf companion is interacting with NGC 5812, and
that the companion is bluer than the host galaxy by $(C-R)\sim0.4$ mag, no
indication of the tidal tails can be seen in the colour map. We attribute this
lack of blue excess in the colour map to our photometry being too shallow for
the colour of the low density tidal tails to be measured, or that it is simply
overwhelmed by the much more luminous body of NGC 5812.

\subsection{Galaxy and globular cluster age of NGC 3585}

The age quoted in Table~\ref{paramtable} for NGC 3585 is apparently very young
\cite[$\sim1.72$\,Gyr --][]{Michard06}. We felt it was important to discuss
this, because it seems inconsistent with the colour we have calculated for
this galaxy, there are no obvious merger remnants and there is no obvious
colour gradient, in contrast to what is expected of young galaxies. To address
this we now look at several issues, namely the GCLF and D-$\sigma$
\cite[][]{Dressler87} distances, the GCLF turnover magnitude and the blue
objects $(C-R)\sim0.5$ shown in Figure~\ref{GChist}.

Firstly, to verify our colour of $(C-R)\sim1.7$ is correct, we compared this
with the theoretical conversion to $(B-I)$ by \cite{Richtler12}. We find that
the $(B-I)$ colour given by \cite{Prugniel98} is in good agreement with our
$(C-R)$ colour.

The only measured GCLF distance to NGC 3585 is that by \cite{Humphrey09} of
$\sim24$\,Mpc (M$_{V,TO}\sim-6.75$), which is in stark contrast to the
D$_n$-$\sigma$ distance estimates by \cite{Willick97} of
$\sim12.9-15$\,Mpc. The magnitude of the GCLF turnover by \cite{Humphrey09},
in comparison to the 1 mag brighter ``Universal'' M$_{V,TO}\sim-7.4$ for {\it
  old} GCs \cite[][]{Rejkuba12}, indicates that this GC population contains a
young component, because young populations introduce a bias towards fainter
turnover magnitudes. Furthermore, the large difference between the GCLF and
D$_n$-$\sigma$ distances is also indicative of a young component in the GC
population, as well as a younger stellar population of the galaxy itself. This
is because a fainter GCLF turnover magnitude implies greater distance, whereas
a younger, and, therefore, brighter galactic stellar population will bias the
distance measurement to smaller D$_n$-$\sigma$ distances. Therefore, a
deviation between the D$_n$-$\sigma$ and GCLF distance estimates points to a
young galaxy \cite[see][for a detailed discussion]{Richtler03}. We conclude,
therefore, that both NGC 3585 itself and its GC population must contain
youthful components. Furthermore, in the case of the GC population, we point
to the blue, $(C-R)\sim0.5$, peak in the GC distribution (Figure~\ref{GChist})
as being this young population.

It is interesting that the $(C-R)$ colour of the galaxy light does not
indicate a young galactic stellar population. Nor is there any colour gradient
in the galactic stellar body, or any obvious tidal debris associated with this
galaxy, which would confirm its youthfulness. We cannot conclude, therefore,
anything with regard to the young age of the galactic population itself,
except to state that its apparent immaturity remains a mystery, at least for
the time being.

\subsection{Globular cluster systems}\label{GCsystems}

The globular cluster systems of each galaxy have been characterised here for
the first time in the Washington photometric system, with $\sim130$ and
$\sim180$ GC candidates being uncovered for NGC 3585 and NGC 5812,
respectively. We find that the radial density profile of both GC systems
closely match a S\'ersic distribution. Assuming the GCLFs follow simple
Gaussian distributions, we calculate that NGC 3585 has $\sim550$ associated
GCs, and for NGC 5812 about 400. The specific frequencies for the two galaxies
are calculated to be $S_N\sim1.05$ and $S_N\sim1.2$ for NGC 3585 and NGC 5812,
respectively. Both galaxies, therefore, have quite poor GC systems, however,
this is typical for field ellipticals. The total number of GC candidates
detected in the current study for NGC 3585 increases the number of known GC
candidates in this galaxy by $\sim40$ from the \cite{Humphrey09} study, and
our estimate of the specific frequency is more than double that of
Humphrey. We attribute these increases to the wide-field nature of our survey,
however, even with these updated values of the total number of GCs and the
specific frequency, NGC 3585 still has a poor GC system, which is expected for
an IE. Both GC systems exhibit the apparently Universal blue peak at
$(C-R)\sim1.3$ ($\sigma_{\rm 3585}\sim0.28$ and $\sigma_{\rm 5812}\sim0.18$),
and both systems exhibit clear colour bimodality, with the red peaks at
$(C-R)_{\rm 3585}\sim1.91$ ($\sigma\sim0.14$) and $(C-R)_{\rm 5812}\sim1.68$
($\sigma\sim0.07$).

There appears to be a Universal blue peak in old GC populations of early-type
galaxies \cite[$[C-T_1{]}\sim1.3$: e.g.][and the current
  paper]{Dirsch03b,Dirsch05,Lee08,Park10,Schuberth10,Richtler12} which likely
indicates that the merger histories of ellipticals are essentially ``dry''
(gas poor) and dominated by old stellar populations \cite[see also][for
  further discussion]{Tal09}. Interestingly, there are hints that this
Universal colour peak also extends to nearby spiral galaxies \cite[e.g.][who
  find that the peak colour of the blue GC population of M31 is at
  $[C-T_1{]}\sim1.4$]{Kim07}, however, this does not seem to be true for the
Milky Way \cite[][]{Smith07}.

The lack of obvious star formation due to the dwarf interacting with NGC 5812
seems to imply that this particular interaction is also fairly dry, at least
at the present time. However, there is clearly an excess of objects bluer than
the blue peak at $0\lesssim(C-R)\lesssim0.9$. It seems that these objects can
come from two sources. Firstly, accretion of young clusters from the
interacting dwarf. We do not consider this the most likely scenario because
there is no reason for the dwarf to contain GCs which are younger, i.e. bluer,
than the dwarf itself ($[C-R]_{\rm dwarf}\sim1.3$). Secondly, \cite{Proctor05}
showed that a star burst with $\sim0.2\%$ of the larger galaxy mass can be
triggered solely by the tidal forces exerted by a low mass encounter. If we
assume that all the GCs have been produced by the interaction and are now
evenly distributed about the host (Section \ref{GCsection}), we can estimate
the number of GCs that have been formed. If we take the maximum/minimum
distances given by NED, m$_R=16.6$ (Section \ref{5812dwarf}) and $(V-R)=0.66$
(Table \ref{paramtable}), the dwarf companion has
$14.5\lesssim$~M$_V\lesssim15.2$. The specific frequency of dwarf spheroidal
galaxies with the absolute magnitude range of the dwarf companion of NGC 5812
is $2\lesssim S_N\lesssim30$
\cite[e.g.][]{vandenBergh98,Elmegreen99,Miller07}. The total number of GCs in
the system may be, therefore, as few as $\sim2$ or as many as $\sim40$. This
means that the blue excess of objects in the NGC 5812 field can be completely
attributed to young, blue, GCs formed in the interaction, since there are 42
objects in the field with $(C-R)<0.9$. It is this scenario which we find much
more satisfactory; the blue GCs formed, and may still be forming, due to
dynamical processes from the interaction. The most likely process being tidal
influences, because of the apparent dryness of the merger. Not only does this
scenario explain the excess of blue objects associated with NGC 5812, but
explains the high specific frequency of the companion ($\gtrsim30$). Moreover,
it is apparent from Figure \ref{GChist} that the blue GC candidates in
question are evenly distributed in colour. This colour distibution is expected
if these GCs have been formed at a constant rate (assuming similar cluster
masses, metallicities, etc), which seems likely if they have been produced by
the interaction between NGC 5812 and its dwarf companion. In addition, the
blue GC candidates are evenly distibuted about NGC 5812, with no obvious
concentrations near the dwarf companion, or other tidal features. This should
be expected because it is clear that the dwarf has been interacting with NGC
5812 for several orbits, based on the length of the tidal tails and the
abudance of other tidal debris surrounding NGC 5812.

There appear to be a few ($\lesssim5$) GC candidates associated with each
galaxy that are very bright, with ${\rm M}_R\sim-13$ (see Figure~\ref{lmftn}).
These clusters must be massive, however, no {\it HST} data appears to exist
for these objects (Section~\ref{3585GCs}). This indicates that they were not
formed in a recent starburst, which would be expected near the centre of the
galaxy, but have been accreted.

Lastly, it is quite clear that the suppressed blue peak of the old GC
population of NGC 3585 is a product of its isolated environment. The isolation
of this galaxy can also be gleaned from its very low $S_N$. Because the old
GCs do not dominate the GC population, as they do in NGC 5812, this is highly
indicative that the host galaxy environment plays a role in shaping its GC
system.

\begin{acknowledgements}
RRL acknowledges financial support from FONDECYT, project No. 3130403. TR
acknowledges financial support from the Chilean Center for Astrophysics,
FONDAP No. 15010003, from FONDECYT project No. 1100620, and from the BASAL
Centro de Astrofisica y Tecnologias Afines (CATA) PFB-06/2007. RS thanks Paula
Zelaya for interesting discussions. The authors gratefully thank the anonymous
referee for many helpful comments which resulted in a much improved
manuscript.

\end{acknowledgements}

\begin{appendix}
\section{IRAF {\it ellipse} model tables}

\onecolumn

\begin{center}
\begin{longtable}{ccccccccccccc}
\caption{{\it ellipse} model of NGC 3585. Columns are: 1 - semi-major axis in arcseconds; 2 - mean isolphotal intensity; 3 - uncertainty in mean isophotal intensity (RMS/sqrt(NDATA)); 4 - ellipticity; 5 - ellipticity error; 6 - position angle (degrees anticlokwise from $+y$); 7 - position angle error; 8 - total flux enclosed by isophote; 9 - 3rd harmonic deviation from a pure ellipse; 10 - uncertaintly in A3; 11 - 4th harmonic deviation from a pure ellipse; 12 - uncertaintly in A4}\\
\hline
SMA & INTENS & INT\_E & ELLIP & ELLIP\_E & PA & PA\_E & TFL\_ELL & A3 & A3\_E & A4 & A4\_E \\
\hline
\endhead
0.00 & \#\# & \#\# & \#\# & \#\# & \#\# & \#\# & \#\# & \#\# & \#\# & \#\# & \#\# \\
0.14 & 636.000 & 0.152 & 0.425 & 0.000 & 35.67 & 0.00 & 635 & -8.53e+00 & 2.43e+01 & -1.38e+00 & 2.32e+01 \\
0.15 & 636.000 & 0.167 & 0.425 & 0.000 & 35.67 & 0.00 & 635 & -7.43e+00 & 1.97e+01 & -3.17e-01 & 1.87e+01 \\
0.17 & 636.000 & 0.189 & 0.425 & 0.000 & 35.67 & 0.00 & 635 & -6.56e+00 & 1.60e+01 & 2.79e-01 & 1.50e+01 \\
0.19 & 636.000 & 0.207 & 0.425 & 0.000 & 35.67 & 0.00 & 635 & -5.75e+00 & 1.30e+01 & 6.01e-01 & 1.21e+01 \\
0.21 & 636.000 & 0.234 & 0.425 & 0.000 & 35.67 & 0.00 & 635 & -4.98e+00 & 1.06e+01 & 7.79e-01 & 9.76e+00 \\
0.23 & 636.000 & 0.254 & 0.425 & 0.000 & 35.67 & 0.00 & 635 & -4.15e+00 & 8.55e+00 & 6.74e-01 & 7.85e+00 \\
0.25 & 636.000 & 0.287 & 0.425 & 0.000 & 35.67 & 0.00 & 635 & -3.57e+00 & 6.95e+00 & 4.61e-01 & 6.29e+00 \\
0.28 & 636.000 & 0.314 & 0.425 & 0.000 & 35.67 & 0.00 & 635 & -3.07e+00 & 5.66e+00 & 2.93e-01 & 5.04e+00 \\
0.30 & 636.000 & 0.345 & 0.425 & 0.000 & 35.67 & 0.00 & 635 & -2.61e+00 & 4.60e+00 & 1.04e-01 & 4.02e+00 \\
0.33 & 636.000 & 0.379 & 0.425 & 0.000 & 35.67 & 0.00 & 635 & -1.98e+00 & 3.65e+00 & -1.55e-01 & 3.20e+00 \\
0.36 & 637.000 & 0.406 & 0.425 & 0.000 & 35.67 & 0.00 & 1908 & -1.28e+00 & 2.78e+00 & -4.15e-01 & 2.52e+00 \\
0.40 & 637.000 & 0.444 & 0.425 & 0.000 & 35.67 & 0.00 & 3186 & -8.99e-01 & 2.17e+00 & -5.59e-01 & 2.02e+00 \\
0.44 & 637.000 & 0.483 & 0.425 & 0.000 & 35.67 & 0.00 & 4459 & -5.73e-01 & 1.69e+00 & -5.82e-01 & 1.62e+00 \\
0.49 & 637.000 & 0.517 & 0.425 & 0.000 & 35.67 & 0.00 & 4459 & -3.70e-01 & 1.30e+00 & -4.38e-01 & 1.25e+00 \\
0.53 & 637.000 & 0.547 & 0.425 & 0.000 & 35.67 & 0.00 & 4459 & -2.28e-01 & 9.90e-01 & -2.74e-01 & 9.51e-01 \\
0.59 & 638.000 & 0.584 & 0.425 & 0.000 & 35.67 & 0.00 & 4459 & -1.75e-01 & 7.66e-01 & -3.83e-01 & 7.71e-01 \\
0.65 & 638.000 & 0.534 & 0.425 & 0.000 & 35.67 & 0.00 & 5741 & -9.58e-02 & 5.02e-01 & -2.13e-01 & 5.12e-01 \\
0.71 & 638.000 & 0.528 & 0.425 & 0.000 & 35.67 & 0.00 & 8289 & -1.43e-01 & 3.79e-01 & -7.81e-02 & 3.63e-01 \\
0.78 & 638.000 & 0.502 & 0.425 & 0.000 & 35.67 & 0.00 & 9570 & -1.97e-01 & 2.88e-01 & -2.64e-01 & 3.09e-01 \\
0.86 & 638.000 & 0.630 & 0.425 & 0.000 & 35.67 & 0.00 & 12118 & -2.01e-01 & 2.94e-01 & -2.29e-01 & 3.08e-01 \\
0.95 & 638.000 & 0.790 & 0.425 & 0.000 & 35.67 & 0.00 & 13399 & -1.18e-01 & 1.41e-01 & -1.39e-01 & 1.53e-01 \\
1.04 & 637.000 & 0.658 & 0.425 & 0.045 & 35.67 & 3.99 & 15954 & -1.33e-02 & 3.58e-02 & -2.38e-02 & 3.62e-02 \\
1.14 & 636.000 & 0.462 & 0.242 & 0.041 & 68.75 & 5.48 & 28679 & 1.38e-02 & 2.59e-02 & 1.42e-02 & 1.97e-02 \\
1.26 & 636.000 & 0.477 & 0.370 & 0.047 & 53.34 & 4.57 & 28684 & -3.05e-02 & 3.96e-02 & -2.85e-02 & 3.75e-02 \\
1.39 & 637.000 & 0.595 & 0.412 & 0.042 & 67.92 & 3.87 & 29953 & -1.65e-02 & 3.62e-02 & -3.91e-03 & 3.49e-02 \\
1.52 & 636.000 & 0.476 & 0.519 & 0.028 & 63.49 & 2.17 & 29952 & -2.53e-02 & 3.06e-02 & 4.46e-02 & 3.43e-02 \\
1.68 & 637.000 & 0.431 & 0.839 & 0.027 & 67.71 & 1.86 & 13394 & 4.12e-02 & 7.24e-02 & -9.49e-02 & 9.42e-02 \\
1.84 & 637.000 & 0.373 & 0.597 & 0.040 & 75.16 & 2.94 & 37585 & 3.27e-02 & 5.62e-02 & 6.91e-03 & 4.92e-02 \\
2.03 & 635.000 & 0.671 & 0.403 & 0.015 & 83.99 & 1.37 & 70681 & -2.31e-02 & 1.28e-02 & 9.91e-03 & 1.01e-02 \\
2.23 & 635.000 & 0.599 & 0.403 & 0.009 & -83.23 & 0.83 & 79570 & 3.84e-03 & 7.48e-03 & -2.88e-02 & 6.91e-03 \\
2.45 & 630.000 & 1.250 & 0.435 & 0.009 & -84.25 & 0.82 & 92317 & 6.91e-03 & 8.08e-03 & -2.64e-02 & 6.73e-03 \\
2.70 & 614.000 & 2.750 & 0.451 & 0.011 & -84.86 & 0.99 & 108614 & 3.86e-03 & 1.05e-02 & -3.22e-02 & 8.99e-03 \\
2.97 & 573.000 & 5.580 & 0.446 & 0.017 & -84.14 & 1.40 & 132396 & 5.99e-03 & 1.48e-02 & -4.97e-02 & 1.32e-02 \\
3.27 & 523.000 & 7.870 & 0.452 & 0.020 & -83.90 & 1.70 & 152505 & 4.63e-03 & 1.83e-02 & -5.19e-02 & 1.67e-02 \\
3.59 & 452.000 & 6.980 & 0.433 & 0.020 & -82.75 & 1.76 & 187643 & 4.64e-03 & 1.78e-02 & -4.75e-02 & 1.70e-02 \\
3.95 & 386.000 & 5.440 & 0.405 & 0.020 & -80.40 & 1.83 & 219120 & 8.29e-03 & 1.69e-02 & -3.67e-02 & 1.64e-02 \\
4.35 & 333.000 & 3.690 & 0.386 & 0.018 & -78.05 & 1.66 & 258415 & 1.88e-02 & 1.43e-02 & -1.98e-02 & 1.38e-02 \\
4.78 & 292.000 & 0.522 & 0.380 & 0.003 & -75.68 & 0.30 & 290898 & -3.29e-04 & 2.58e-03 & 2.11e-03 & 1.37e-03 \\
5.26 & 271.000 & 0.551 & 0.404 & 0.003 & -75.73 & 0.31 & 320916 & 2.24e-03 & 2.87e-03 & -7.08e-04 & 1.52e-03 \\
5.79 & 249.000 & 0.565 & 0.424 & 0.004 & -75.58 & 0.31 & 350968 & 3.47e-03 & 3.08e-03 & -1.55e-04 & 1.44e-03 \\
6.37 & 229.000 & 0.612 & 0.443 & 0.004 & -75.39 & 0.33 & 384067 & 4.78e-03 & 3.45e-03 & -6.49e-05 & 1.72e-03 \\
7.00 & 209.000 & 0.613 & 0.461 & 0.004 & -75.37 & 0.35 & 421389 & 4.57e-03 & 3.90e-03 & -4.82e-04 & 2.00e-03 \\
7.70 & 189.000 & 0.639 & 0.474 & 0.004 & -75.20 & 0.35 & 462829 & 4.80e-03 & 4.14e-03 & 3.18e-04 & 2.31e-03 \\
8.47 & 170.000 & 0.620 & 0.482 & 0.004 & -75.32 & 0.35 & 508281 & 2.09e-03 & 4.26e-03 & 8.83e-04 & 2.61e-03 \\
9.32 & 151.000 & 0.625 & 0.488 & 0.005 & -75.20 & 0.38 & 558694 & 2.85e-03 & 4.63e-03 & -4.35e-04 & 3.16e-03 \\
10.25 & 132.000 & 0.593 & 0.489 & 0.005 & -75.25 & 0.39 & 615167 & 2.80e-03 & 4.87e-03 & -4.26e-04 & 3.56e-03 \\
11.28 & 114.000 & 0.541 & 0.485 & 0.005 & -75.27 & 0.42 & 677977 & 2.96e-03 & 5.07e-03 & 7.84e-05 & 3.80e-03 \\
12.41 & 97.100 & 0.470 & 0.478 & 0.005 & -75.17 & 0.41 & 744101 & 2.07e-03 & 4.94e-03 & -2.40e-04 & 3.76e-03 \\
13.65 & 82.500 & 0.394 & 0.471 & 0.005 & -75.22 & 0.42 & 815624 & 5.48e-04 & 4.85e-03 & -2.29e-03 & 3.67e-03 \\
15.01 & 70.200 & 0.322 & 0.465 & 0.005 & -75.19 & 0.40 & 888163 & -6.05e-04 & 4.58e-03 & -1.14e-03 & 3.42e-03 \\
16.51 & 59.900 & 0.264 & 0.462 & 0.005 & -75.06 & 0.40 & 961658 & 2.17e-04 & 4.48e-03 & -2.40e-03 & 3.37e-03 \\
18.17 & 51.500 & 0.210 & 0.462 & 0.004 & -75.01 & 0.37 & 1035236 & 1.10e-03 & 4.19e-03 & -1.41e-04 & 3.16e-03 \\
19.98 & 44.000 & 0.164 & 0.458 & 0.004 & -75.03 & 0.34 & 1114492 & 1.62e-03 & 3.77e-03 & 8.93e-04 & 2.85e-03 \\
21.98 & 37.700 & 0.126 & 0.455 & 0.004 & -75.14 & 0.32 & 1196988 & 5.59e-04 & 3.57e-03 & -1.48e-04 & 2.67e-03 \\
24.18 & 32.600 & 0.097 & 0.453 & 0.003 & -75.38 & 0.28 & 1282436 & -1.91e-03 & 3.12e-03 & -9.25e-04 & 2.44e-03 \\
26.60 & 28.000 & 0.071 & 0.449 & 0.003 & -75.27 & 0.25 & 1373801 & -4.26e-04 & 2.67e-03 & -3.78e-04 & 2.01e-03 \\
29.25 & 24.000 & 0.049 & 0.445 & 0.002 & -74.91 & 0.20 & 1469313 & 1.81e-03 & 2.17e-03 & -8.40e-04 & 1.56e-03 \\
32.18 & 20.700 & 0.040 & 0.438 & 0.002 & -74.56 & 0.20 & 1571619 & -1.68e-03 & 2.08e-03 & -6.73e-03 & 1.61e-03 \\
35.40 & 17.700 & 0.026 & 0.431 & 0.002 & -74.78 & 0.15 & 1680049 & -1.61e-03 & 1.50e-03 & -2.77e-03 & 1.12e-03 \\
38.94 & 15.000 & 0.018 & 0.419 & 0.002 & -74.56 & 0.13 & 1797784 & 2.77e-04 & 1.26e-03 & -1.54e-03 & 1.04e-03 \\
42.83 & 12.800 & 0.013 & 0.409 & 0.001 & -74.77 & 0.10 & 1920688 & -2.69e-03 & 9.81e-04 & -2.70e-03 & 8.62e-04 \\
47.11 & 10.600 & 0.008 & 0.393 & 0.001 & -74.87 & 0.07 & 2051566 & -2.30e-03 & 6.58e-04 & 1.19e-04 & 5.30e-04 \\
51.82 & 8.640 & 0.009 & 0.370 & 0.001 & -74.52 & 0.12 & 2193360 & 4.13e-03 & 9.39e-04 & 9.16e-03 & 8.81e-04 \\
57.01 & 7.550 & 0.007 & 0.381 & 0.001 & -75.41 & 0.10 & 2304593 & -1.11e-02 & 7.76e-04 & -4.03e-03 & 6.97e-04 \\
62.71 & 6.410 & 0.011 & 0.381 & 0.002 & -76.14 & 0.18 & 2432535 & -1.50e-02 & 1.48e-03 & -4.51e-03 & 1.45e-03 \\
68.98 & 5.290 & 0.005 & 0.373 & 0.001 & -74.87 & 0.09 & 2573213 & -5.80e-03 & 7.13e-04 & 7.67e-04 & 7.04e-04 \\
75.88 & 4.410 & 0.003 & 0.374 & 0.001 & -74.91 & 0.07 & 2702806 & -9.92e-03 & 5.48e-04 & 8.81e-05 & 5.39e-04 \\
83.47 & 3.720 & 0.003 & 0.377 & 0.001 & -74.73 & 0.09 & 2830739 & -9.68e-03 & 7.11e-04 & -3.03e-03 & 7.04e-04 \\
91.81 & 3.130 & 0.002 & 0.383 & 0.001 & -74.83 & 0.07 & 2958537 & -1.08e-02 & 5.94e-04 & 1.41e-03 & 5.73e-04 \\
100.99 & 2.620 & 0.002 & 0.389 & 0.001 & -74.91 & 0.09 & 3086264 & -1.51e-02 & 6.96e-04 & 1.09e-03 & 6.73e-04 \\
111.09 & 2.200 & 0.002 & 0.395 & 0.001 & -75.19 & 0.08 & 3215381 & -1.14e-02 & 7.12e-04 & -3.51e-03 & 6.91e-04 \\
122.20 & 1.830 & 0.002 & 0.403 & 0.001 & -75.35 & 0.09 & 3380665 & -1.67e-02 & 8.09e-04 & -2.30e-03 & 7.75e-04 \\
134.42 & 1.510 & 0.002 & 0.405 & 0.001 & -75.50 & 0.10 & 3526503 & -1.57e-02 & 9.14e-04 & -4.20e-03 & 8.78e-04 \\
147.86 & 1.230 & 0.002 & 0.405 & 0.001 & -74.99 & 0.10 & 3675320 & -8.34e-03 & 9.69e-04 & -3.49e-03 & 9.59e-04 \\
162.65 & 1.010 & 0.001 & 0.412 & 0.001 & -74.61 & 0.11 & 3811083 & -1.86e-02 & 1.04e-03 & -6.13e-03 & 9.86e-04 \\
178.91 & 0.818 & 0.001 & 0.412 & 0.001 & -75.33 & 0.13 & 3962040 & -1.03e-02 & 1.22e-03 & -8.01e-03 & 1.16e-03 \\
196.81 & 0.688 & 0.001 & 0.426 & 0.002 & -76.49 & 0.14 & 4094684 & 5.41e-04 & 1.44e-03 & -1.01e-03 & 1.38e-03 \\
216.49 & 0.554 & 0.001 & 0.424 & 0.002 & -75.24 & 0.15 & 4288675 & 8.60e-03 & 1.52e-03 & -6.23e-03 & 1.49e-03 \\
238.13 & 0.418 & 0.001 & 0.415 & 0.001 & -74.63 & 0.13 & 4459784 & 5.92e-03 & 1.23e-03 & -8.13e-03 & 1.20e-03 \\
261.95 & 0.336 & 0.001 & 0.419 & 0.002 & -72.26 & 0.17 & 4594577 & 2.75e-03 & 1.72e-03 & -5.94e-03 & 1.72e-03 \\
288.14 & 0.252 & 0.001 & 0.412 & 0.002 & -70.16 & 0.16 & 4800169 & 3.78e-03 & 1.51e-03 & -1.93e-02 & 1.49e-03 \\
316.96 & 0.154 & 0.001 & 0.375 & 0.003 & -70.13 & 0.24 & 5022625 & 3.87e-03 & 1.99e-03 & 2.25e-03 & 1.99e-03 \\
348.65 & 0.100 & 0.001 & 0.375 & 0.003 & -70.13 & 0.26 & 5202617 & -7.16e-03 & 2.18e-03 & -1.48e-02 & 2.19e-03 \\
383.52 & 0.062 & 0.001 & 0.375 & 0.003 & -72.70 & 0.32 & 5437676 & -3.89e-02 & 2.91e-03 & 5.49e-04 & 2.64e-03 \\
\hline
\label{apptable1}\\
\end{longtable}
\end{center}

\begin{center}
\begin{longtable}{ccccccccccccc}
\caption{The same as Table~\ref{apptable1}, except for NGC 5812}\tabularnewline
\hline
SMA & INTENS & INT{\_}E & ELLIP & ELLIP{\_}E & PA & PA{\_}E & TFL{\_}ELL & A3 & A3{\_}E & A4 & A4{\_}E \\
\hline
\endhead
0.00 & \#\# & \#\# & \#\# & \#\# & \#\# & \#\# & \#\# & \#\# & \#\# & \#\# & \#\# \\
0.14 & 763.000 & 0.064 & 0.844 & 0.018 & -1.72 & 1.10 & 764 & -9.05e-03 & 4.92e-02 & -3.40e-02 & 5.41e-02 \\
0.15 & 762.000 & 0.115 & 0.809 & 0.026 & -1.78 & 1.62 & 764 & -6.43e-03 & 5.71e-02 & -4.35e-02 & 6.37e-02 \\
0.17 & 761.000 & 0.235 & 0.638 & 0.049 & -1.78 & 3.31 & 764 & 2.98e-02 & 4.81e-02 & -1.09e-02 & 4.49e-02 \\
0.19 & 759.000 & 0.352 & 0.423 & 0.076 & -6.05 & 6.75 & 764 & -1.35e-02 & 2.87e-02 & -2.27e-02 & 2.96e-02 \\
0.21 & 757.000 & 0.407 & 0.277 & 0.079 & -16.73 & 9.82 & 764 & -8.64e-02 & 4.26e-02 & -1.52e-02 & 1.74e-02 \\
0.23 & 756.000 & 0.441 & 0.225 & 0.087 & -29.38 & 12.98 & 764 & -1.57e-01 & 8.03e-02 & 1.42e-02 & 1.18e-02 \\
0.25 & 756.000 & 0.499 & 0.279 & 0.106 & -37.70 & 13.05 & 764 & -1.93e-01 & 1.29e-01 & 4.85e-02 & 3.59e-02 \\
0.28 & 756.000 & 0.698 & 0.351 & 0.117 & -47.74 & 12.03 & 764 & -1.77e-01 & 1.42e-01 & 7.44e-02 & 6.41e-02 \\
0.30 & 757.000 & 0.733 & 0.527 & 0.156 & -62.60 & 11.88 & 764 & -1.28e-01 & 2.14e-01 & 3.01e-02 & 9.43e-02 \\
0.33 & 758.000 & 0.756 & 0.667 & 0.144 & -68.56 & 9.63 & 758 & -1.53e-01 & 3.39e-01 & 5.66e-02 & 1.86e-01 \\
0.36 & 758.000 & 0.665 & 0.752 & 0.158 & -68.56 & 9.97 & 758 & -2.32e-01 & 7.59e-01 & 4.65e-02 & 2.74e-01 \\
0.40 & 758.000 & 1.020 & 0.741 & 0.122 & -68.56 & 7.76 & 758 & -1.57e-01 & 3.81e-01 & 7.28e-02 & 2.27e-01 \\
0.44 & 757.000 & 1.100 & 0.767 & 0.118 & -70.49 & 7.38 & 758 & -1.90e-01 & 4.99e-01 & 4.27e-02 & 2.10e-01 \\
0.49 & 757.000 & 0.960 & 0.782 & 0.112 & -76.89 & 6.94 & 2285 & -2.77e-01 & 7.24e-01 & -1.01e-02 & 1.96e-01 \\
0.53 & 752.000 & 1.020 & 0.520 & 0.046 & -79.11 & 3.57 & 2285 & -1.00e-01 & 6.56e-02 & -2.77e-02 & 3.31e-02 \\
0.59 & 746.000 & 1.160 & 0.435 & 0.032 & -85.74 & 2.74 & 8312 & -5.35e-02 & 2.56e-02 & -4.14e-02 & 1.90e-02 \\
0.65 & 734.000 & 0.934 & 0.354 & 0.017 & -88.79 & 1.74 & 8312 & -1.39e-02 & 1.12e-02 & -3.07e-02 & 8.64e-03 \\
0.71 & 710.000 & 1.760 & 0.234 & 0.025 & 88.22 & 3.58 & 12680 & 1.32e-02 & 1.35e-02 & -3.17e-02 & 1.09e-02 \\
0.78 & 687.000 & 2.150 & 0.203 & 0.021 & 86.50 & 3.40 & 15499 & 1.06e-02 & 1.05e-02 & -2.49e-02 & 6.12e-03 \\
0.86 & 656.000 & 2.170 & 0.173 & 0.018 & 86.02 & 3.31 & 19496 & 6.64e-03 & 9.30e-03 & -2.34e-02 & 5.37e-03 \\
0.95 & 616.000 & 1.910 & 0.135 & 0.014 & 87.43 & 3.22 & 23377 & -1.46e-03 & 7.23e-03 & -2.08e-02 & 4.20e-03 \\
1.04 & 574.000 & 1.630 & 0.112 & 0.011 & -89.73 & 3.14 & 28138 & -4.05e-03 & 5.75e-03 & -1.74e-02 & 3.75e-03 \\
1.14 & 528.000 & 1.030 & 0.080 & 0.007 & -84.95 & 2.77 & 33682 & -6.24e-03 & 3.47e-03 & -1.17e-02 & 2.46e-03 \\
1.26 & 485.000 & 0.360 & 0.063 & 0.003 & -80.99 & 1.23 & 40795 & 2.57e-04 & 1.34e-03 & -3.60e-03 & 1.11e-03 \\
1.39 & 443.000 & 0.360 & 0.050 & 0.003 & -78.43 & 1.62 & 46307 & 3.05e-03 & 1.29e-03 & 2.53e-03 & 1.20e-03 \\
1.52 & 405.000 & 0.229 & 0.049 & 0.002 & -79.67 & 1.08 & 53999 & 2.67e-03 & 7.92e-04 & 1.50e-03 & 7.29e-04 \\
1.68 & 368.000 & 0.175 & 0.043 & 0.001 & -79.35 & 0.97 & 61707 & 1.00e-03 & 7.05e-04 & -7.01e-04 & 6.91e-04 \\
1.84 & 332.000 & 0.175 & 0.037 & 0.002 & -77.58 & 1.21 & 70802 & 2.59e-03 & 6.09e-04 & -4.77e-04 & 6.03e-04 \\
2.03 & 298.000 & 0.145 & 0.033 & 0.001 & -75.14 & 1.24 & 81405 & 2.52e-03 & 5.84e-04 & -2.03e-04 & 5.34e-04 \\
2.23 & 268.000 & 0.132 & 0.029 & 0.001 & -69.41 & 1.35 & 91431 & 1.54e-03 & 6.43e-04 & 4.29e-04 & 5.97e-04 \\
2.45 & 240.000 & 0.132 & 0.029 & 0.002 & -61.35 & 1.46 & 100595 & 2.51e-03 & 6.62e-04 & 1.91e-03 & 6.07e-04 \\
2.70 & 213.000 & 0.262 & 0.015 & 0.002 & -70.00 & 3.68 & 113378 & 2.55e-03 & 8.85e-04 & 3.07e-03 & 7.79e-04 \\
2.97 & 190.000 & 0.226 & 0.015 & 0.002 & -70.00 & 3.48 & 127549 & 1.30e-03 & 8.73e-04 & 3.18e-03 & 6.31e-04 \\
3.27 & 170.000 & 0.238 & 0.015 & 0.002 & -70.00 & 3.00 & 141237 & -2.40e-06 & 7.64e-04 & 2.88e-03 & 4.74e-04 \\
3.59 & 150.000 & 0.241 & 0.015 & 0.002 & -70.00 & 3.04 & 157520 & 9.55e-04 & 7.75e-04 & 3.01e-03 & 5.85e-04 \\
3.95 & 133.000 & 0.230 & 0.015 & 0.002 & -70.00 & 3.22 & 174135 & 7.16e-04 & 8.29e-04 & 3.64e-03 & 5.37e-04 \\
4.35 & 117.000 & 0.216 & 0.015 & 0.002 & -70.00 & 3.35 & 190342 & -6.92e-05 & 8.65e-04 & 4.01e-03 & 5.44e-04 \\
4.78 & 103.000 & 0.167 & 0.015 & 0.002 & -70.00 & 2.99 & 208803 & 3.02e-04 & 7.73e-04 & 3.69e-03 & 4.75e-04 \\
5.26 & 89.000 & 0.111 & 0.015 & 0.001 & -69.01 & 2.64 & 229400 & -5.90e-05 & 6.81e-04 & 3.82e-03 & 4.53e-04 \\
5.79 & 77.100 & 0.044 & 0.020 & 0.001 & -56.91 & 1.62 & 248560 & 4.74e-04 & 5.66e-04 & 3.33e-03 & 4.89e-04 \\
6.37 & 65.800 & 0.041 & 0.015 & 0.001 & -73.54 & 2.44 & 270460 & -1.09e-03 & 6.13e-04 & 3.34e-03 & 5.52e-04 \\
7.00 & 56.500 & 0.035 & 0.019 & 0.001 & 79.06 & 1.80 & 291657 & 1.37e-03 & 5.85e-04 & -3.78e-04 & 5.66e-04 \\
7.70 & 48.600 & 0.032 & 0.027 & 0.001 & 75.68 & 1.38 & 313632 & 4.92e-04 & 6.34e-04 & 2.69e-04 & 6.34e-04 \\
8.47 & 41.900 & 0.031 & 0.033 & 0.001 & 70.61 & 1.22 & 336218 & 1.60e-03 & 6.93e-04 & 2.55e-03 & 6.68e-04 \\
9.32 & 35.700 & 0.027 & 0.028 & 0.001 & 60.11 & 1.44 & 361199 & 2.13e-03 & 6.68e-04 & 2.52e-03 & 6.42e-04 \\
10.25 & 30.300 & 0.017 & 0.032 & 0.001 & 54.64 & 0.89 & 385824 & 2.28e-03 & 4.67e-04 & 1.15e-04 & 4.66e-04 \\
11.28 & 25.700 & 0.016 & 0.036 & 0.001 & 54.10 & 0.87 & 410512 & 3.09e-03 & 5.25e-04 & -4.06e-04 & 5.22e-04 \\
12.41 & 21.600 & 0.015 & 0.039 & 0.001 & 51.55 & 0.94 & 436395 & 2.84e-03 & 6.17e-04 & 8.48e-04 & 5.75e-04 \\
13.65 & 18.300 & 0.012 & 0.042 & 0.001 & 53.24 & 0.82 & 462646 & 4.31e-04 & 6.01e-04 & -2.31e-03 & 5.39e-04 \\
15.01 & 15.500 & 0.012 & 0.044 & 0.001 & 56.29 & 0.92 & 489252 & 1.57e-03 & 6.95e-04 & -4.24e-03 & 6.24e-04 \\
16.51 & 13.200 & 0.013 & 0.051 & 0.002 & 57.05 & 0.98 & 516118 & 2.42e-03 & 8.79e-04 & -6.03e-03 & 7.54e-04 \\
18.17 & 11.100 & 0.011 & 0.049 & 0.002 & 56.08 & 0.98 & 544939 & 1.92e-03 & 8.27e-04 & -4.51e-03 & 7.27e-04 \\
19.98 & 9.370 & 0.008 & 0.052 & 0.002 & 55.62 & 0.85 & 573512 & 1.72e-03 & 7.74e-04 & -2.45e-03 & 6.58e-04 \\
21.98 & 7.910 & 0.006 & 0.051 & 0.001 & 54.09 & 0.76 & 602966 & 2.37e-03 & 6.88e-04 & -1.33e-03 & 6.26e-04 \\
24.18 & 6.740 & 0.005 & 0.056 & 0.001 & 57.53 & 0.66 & 632497 & -1.91e-04 & 6.56e-04 & -1.73e-03 & 6.36e-04 \\
26.60 & 5.770 & 0.005 & 0.060 & 0.002 & 58.51 & 0.74 & 662897 & 1.08e-03 & 7.90e-04 & -2.84e-03 & 7.79e-04 \\
29.25 & 4.920 & 0.004 & 0.063 & 0.002 & 57.62 & 0.72 & 694412 & 2.84e-03 & 7.85e-04 & -5.86e-03 & 7.28e-04 \\
32.18 & 4.210 & 0.004 & 0.068 & 0.002 & 58.06 & 0.66 & 726417 & 3.28e-03 & 7.81e-04 & -2.94e-03 & 7.52e-04 \\
35.40 & 3.570 & 0.002 & 0.070 & 0.001 & 59.47 & 0.46 & 759908 & -1.24e-03 & 5.75e-04 & -2.39e-03 & 5.66e-04 \\
38.94 & 3.000 & 0.002 & 0.066 & 0.001 & 66.50 & 0.42 & 795160 & 6.30e-04 & 5.07e-04 & -4.92e-03 & 4.79e-04 \\
42.83 & 2.540 & 0.001 & 0.074 & 0.001 & 65.69 & 0.34 & 828955 & 7.27e-04 & 4.47e-04 & -4.87e-03 & 4.14e-04 \\
47.11 & 2.100 & 0.001 & 0.068 & 0.001 & 64.71 & 0.38 & 865896 & 3.19e-04 & 4.54e-04 & -5.16e-03 & 4.28e-04 \\
51.82 & 1.750 & 0.001 & 0.074 & 0.001 & 63.53 & 0.47 & 900615 & -4.08e-04 & 6.28e-04 & -8.99e-03 & 5.63e-04 \\
57.01 & 1.480 & 0.001 & 0.088 & 0.001 & 64.95 & 0.46 & 933660 & -4.10e-03 & 7.15e-04 & -9.17e-03 & 6.49e-04 \\
62.71 & 1.250 & 0.001 & 0.092 & 0.001 & 63.86 & 0.44 & 969430 & -1.62e-03 & 7.20e-04 & -5.41e-03 & 6.80e-04 \\
68.98 & 1.030 & 0.001 & 0.085 & 0.001 & 70.92 & 0.45 & 1008994 & -1.52e-03 & 6.61e-04 & -1.12e-02 & 5.86e-04 \\
75.88 & 0.820 & 0.001 & 0.086 & 0.001 & 70.01 & 0.42 & 1045095 & -3.42e-03 & 6.57e-04 & -9.61e-03 & 5.96e-04 \\
83.47 & 0.643 & 0.001 & 0.088 & 0.002 & 66.90 & 0.61 & 1079514 & 3.38e-04 & 9.44e-04 & -1.02e-02 & 8.82e-04 \\
91.81 & 0.511 & 0.001 & 0.090 & 0.002 & 59.58 & 0.67 & 1112477 & -2.39e-03 & 1.03e-03 & -1.47e-03 & 1.00e-03 \\
100.99 & 0.405 & 0.001 & 0.102 & 0.002 & 55.05 & 0.72 & 1149325 & -1.52e-02 & 1.30e-03 & 1.38e-02 & 1.24e-03 \\
111.09 & 0.295 & 0.001 & 0.089 & 0.004 & 50.19 & 1.26 & 1205022 & -1.13e-02 & 2.06e-03 & 1.81e-02 & 2.05e-03 \\
122.20 & 0.214 & 0.001 & 0.089 & 0.004 & 43.99 & 1.31 & 1239573 & -2.78e-02 & 2.15e-03 & 3.31e-02 & 1.95e-03 \\
134.42 & 0.153 & 0.001 & 0.089 & 0.004 & 29.64 & 1.38 & 1271169 & -5.31e-02 & 2.52e-03 & -1.63e-02 & 1.96e-03 \\
147.86 & 0.116 & 0.001 & 0.132 & 0.005 & 37.16 & 1.11 & 1287618 & -3.56e-02 & 2.85e-03 & -1.37e-02 & 2.48e-03 \\
162.65 & 0.083 & 0.001 & 0.132 & 0.010 & 22.62 & 2.30 & 1316133 & -8.07e-02 & 8.67e-03 & -8.67e-02 & 8.89e-03 \\
178.91 & 0.078 & 0.001 & 0.209 & 0.008 & 25.46 & 1.23 & 1326195 & -9.45e-02 & 7.51e-03 & -7.36e-02 & 6.45e-03 \\
196.81 & 0.067 & 0.001 & 0.276 & 0.006 & 25.46 & 0.77 & 1358317 & -3.84e-02 & 4.81e-03 & -9.59e-02 & 6.68e-03 \\
216.49 & 0.056 & 0.001 & 0.296 & 0.006 & 19.38 & 0.69 & 1398031 & -2.74e-03 & 4.29e-03 & -7.85e-02 & 5.68e-03 \\
238.14 & 0.036 & 0.001 & 0.314 & 0.006 & 15.96 & 0.62 & 1435704 & 1.57e-02 & 4.24e-03 & -4.18e-02 & 4.59e-03 \\
261.95 & 0.017 & 0.000 & 0.314 & 0.006 & 15.96 & 0.66 & 1487091 & 6.24e-03 & 4.44e-03 & -4.83e-03 & 4.42e-03 \\
\hline
\label{apptable2}\\
\end{longtable}
\end{center}

\begin{center}
\begin{longtable}{ccccccccccccc}
\caption{The same as Table~\ref{apptable1}, except for the dwarf galaxy
  companion of NGC 5812}\tabularnewline
\hline
SMA & INTENS & INT{\_}E & ELLIP & ELLIP{\_}E & PA & PA{\_}E & TFL{\_}ELL & A3 & A3{\_}E & A4 & A4{\_}E \\
\hline
\endhead
0.00 & \#\# & \#\# & \#\# & \#\# & \#\# & \#\# & \#\# & \#\# & \#\# & \#\# & \#\# \\
0.14 & 4.100 & 0.002 & 0.322 & 0.059 & -85.15 & 6.55 & 4.1694 & -0.0402 & 0.027 & -0.0155 & 0.022 \\
0.15 & 4.100 & 0.002 & 0.323 & 0.059 & -84.67 & 6.52 & 4.1694 & -0.0405 & 0.0279 & -0.017 & 0.0231 \\
0.15 & 4.100 & 0.002 & 0.325 & 0.058 & -84.27 & 6.37 & 4.1694 & -0.0399 & 0.028 & -0.0175 & 0.0235 \\
0.16 & 4.100 & 0.002 & 0.330 & 0.057 & -83.98 & 6.11 & 4.1694 & -0.0387 & 0.0276 & -0.0173 & 0.0233 \\
0.17 & 4.090 & 0.002 & 0.333 & 0.054 & -83.65 & 5.84 & 4.1694 & -0.0371 & 0.027 & -0.017 & 0.023 \\
0.18 & 4.090 & 0.002 & 0.337 & 0.052 & -83.34 & 5.55 & 4.1694 & -0.0354 & 0.0264 & -0.016 & 0.0226 \\
0.18 & 4.090 & 0.002 & 0.344 & 0.045 & -82.90 & 4.73 & 4.1694 & -0.0303 & 0.0226 & -0.0137 & 0.02 \\
0.19 & 4.080 & 0.002 & 0.322 & 0.041 & -81.58 & 4.53 & 4.1694 & -0.0261 & 0.0229 & -0.00964 & 0.0211 \\
0.20 & 4.080 & 0.002 & 0.284 & 0.040 & -79.81 & 4.88 & 4.1694 & -0.0185 & 0.0245 & 1.04E-4 & 0.0234 \\
0.21 & 4.070 & 0.002 & 0.261 & 0.042 & -77.81 & 5.47 & 4.1694 & -0.0074 & 0.0263 & 0.00189 & 0.0256 \\
0.22 & 4.060 & 0.002 & 0.241 & 0.048 & -75.47 & 6.72 & 4.1694 & 0.00459 & 0.0277 & 0.00191 & 0.0271 \\
0.23 & 4.050 & 0.002 & 0.193 & 0.045 & -68.72 & 7.74 & 4.1694 & 0.0309 & 0.0214 & 0.014 & 0.0191 \\
0.25 & 4.050 & 0.002 & 0.147 & 0.049 & -63.12 & 10.72 & 4.1694 & 0.0501 & 0.0208 & 0.024 & 0.0143 \\
0.26 & 4.040 & 0.003 & 0.130 & 0.048 & -59.29 & 11.69 & 4.1694 & 0.0534 & 0.0208 & 0.0234 & 0.015 \\
0.27 & 4.030 & 0.003 & 0.112 & 0.045 & -56.56 & 12.86 & 4.1694 & 0.0532 & 0.0203 & 0.016 & 0.0156 \\
0.29 & 4.020 & 0.003 & 0.083 & 0.049 & -53.27 & 18.30 & 20.229 & 0.0599 & 0.0219 & 0.00507 & 0.0157 \\
0.30 & 4.010 & 0.003 & 0.083 & 0.050 & -53.27 & 18.71 & 20.229 & 0.0612 & 0.0224 & 0.00206 & 0.0158 \\
0.32 & 4.000 & 0.004 & 0.060 & 0.056 & -53.27 & 28.42 & 20.229 & 0.0669 & 0.025 & -0.00294 & 0.0163 \\
0.33 & 3.990 & 0.004 & 0.060 & 0.058 & -53.27 & 29.60 & 20.229 & 0.0694 & 0.0258 & -0.00553 & 0.0166 \\
0.35 & 3.970 & 0.004 & 0.046 & 0.055 & -56.00 & 36.32 & 20.229 & 0.068 & 0.0217 & -0.00661 & 0.0144 \\
0.36 & 3.960 & 0.003 & 0.039 & 0.038 & -62.25 & 29.95 & 20.229 & 0.0547 & 0.0113 & -0.00608 & 0.00679 \\
0.38 & 3.950 & 0.003 & 0.039 & 0.033 & -67.42 & 25.67 & 20.229 & 0.0436 & 0.00949 & -0.00815 & 0.00674 \\
0.40 & 3.930 & 0.003 & 0.039 & 0.029 & -67.42 & 23.05 & 36.015 & 0.0383 & 0.00857 & -0.00922 & 0.00658 \\
0.42 & 3.920 & 0.003 & 0.039 & 0.027 & -67.42 & 20.92 & 36.015 & 0.0338 & 0.00801 & -0.00872 & 0.00653 \\
0.44 & 3.900 & 0.003 & 0.032 & 0.024 & -64.53 & 22.89 & 36.015 & 0.033 & 0.00626 & -0.005 & 0.00475 \\
0.47 & 3.880 & 0.003 & 0.032 & 0.023 & -64.53 & 21.66 & 36.015 & 0.0293 & 0.00634 & -0.00545 & 0.0049 \\
0.49 & 3.860 & 0.003 & 0.025 & 0.021 & -61.33 & 25.88 & 36.015 & 0.0265 & 0.0053 & 0.00118 & 0.00384 \\
0.51 & 3.850 & 0.003 & 0.031 & 0.020 & -60.04 & 18.82 & 36.015 & 0.0214 & 0.00594 & 0.00507 & 0.00395 \\
0.54 & 3.820 & 0.003 & 0.031 & 0.020 & -60.04 & 19.26 & 36.015 & 0.0165 & 0.00688 & 0.0066 & 0.00492 \\
0.57 & 3.810 & 0.003 & 0.054 & 0.020 & -74.82 & 11.00 & 43.667 & 0.0152 & 0.00848 & -0.0144 & 0.00413 \\
0.59 & 3.780 & 0.003 & 0.056 & 0.019 & -62.68 & 10.56 & 51.298 & 0.00999 & 0.00828 & 0.0089 & 0.00483 \\
0.62 & 3.760 & 0.004 & 0.056 & 0.024 & -62.68 & 13.49 & 66.431 & 0.00207 & 0.00854 & 0.00245 & 0.00629 \\
0.66 & 3.730 & 0.004 & 0.062 & 0.018 & -63.72 & 8.89 & 81.532 & 0.00309 & 0.00674 & 0.00903 & 0.00575 \\
0.69 & 3.710 & 0.005 & 0.078 & 0.021 & -63.72 & 8.26 & 81.532 & -7.12E-4 & 0.00919 & -0.00496 & 0.00829 \\
0.72 & 3.690 & 0.005 & 0.096 & 0.019 & -68.25 & 6.26 & 81.532 & 0.0122 & 0.00701 & 0.00101 & 0.0066 \\
0.76 & 3.650 & 0.004 & 0.094 & 0.017 & -65.06 & 5.51 & 81.532 & 0.00541 & 0.00586 & 0.00198 & 0.00579 \\
0.80 & 3.620 & 0.004 & 0.107 & 0.014 & -65.99 & 4.25 & 88.884 & 7.81E-4 & 0.0065 & 0.00355 & 0.00639 \\
0.84 & 3.590 & 0.004 & 0.113 & 0.016 & -64.49 & 4.52 & 96.127 & 0.00475 & 0.00763 & -4.63E-4 & 0.00713 \\
0.88 & 3.550 & 0.004 & 0.117 & 0.015 & -63.38 & 4.28 & 110.38 & 0.00565 & 0.00788 & -0.00213 & 0.00695 \\
0.92 & 3.520 & 0.004 & 0.120 & 0.014 & -60.82 & 3.94 & 131.46 & 0.00243 & 0.00782 & -0.00783 & 0.00671 \\
0.97 & 3.490 & 0.004 & 0.120 & 0.014 & -58.96 & 3.73 & 138.5 & 0.00535 & 0.00772 & -0.0115 & 0.00618 \\
1.02 & 3.450 & 0.003 & 0.119 & 0.011 & -55.27 & 2.94 & 152.45 & 0.00494 & 0.00599 & -0.0134 & 0.00495 \\
1.07 & 3.410 & 0.003 & 0.119 & 0.009 & -50.75 & 2.47 & 152.45 & 8.05E-4 & 0.00508 & -0.0124 & 0.00435 \\
1.12 & 3.380 & 0.002 & 0.125 & 0.006 & -47.29 & 1.64 & 166. & -0.00396 & 0.00354 & -0.00765 & 0.00285 \\
1.18 & 3.350 & 0.002 & 0.134 & 0.007 & -44.60 & 1.84 & 199.65 & -0.00407 & 0.00436 & -0.00449 & 0.00381 \\
1.24 & 3.310 & 0.003 & 0.142 & 0.011 & -43.04 & 2.44 & 212.95 & -0.00195 & 0.00628 & -0.00108 & 0.00586 \\
1.30 & 3.280 & 0.003 & 0.153 & 0.011 & -41.65 & 2.33 & 226.13 & 0.00231 & 0.00647 & -0.00297 & 0.00626 \\
1.36 & 3.250 & 0.003 & 0.165 & 0.010 & -39.60 & 1.95 & 232.63 & 0.00577 & 0.00585 & -0.00773 & 0.00548 \\
1.43 & 3.220 & 0.003 & 0.176 & 0.010 & -37.73 & 1.90 & 252.16 & 0.00756 & 0.00609 & -0.01 & 0.00558 \\
1.50 & 3.180 & 0.004 & 0.190 & 0.011 & -35.67 & 1.88 & 277.77 & 0.0101 & 0.00648 & -0.00666 & 0.0061 \\
1.58 & 3.150 & 0.004 & 0.204 & 0.010 & -34.18 & 1.64 & 303.08 & 0.0141 & 0.00569 & -1.06E-5 & 0.00545 \\
1.66 & 3.110 & 0.004 & 0.218 & 0.010 & -32.43 & 1.60 & 315.68 & 0.0151 & 0.00546 & 0.00663 & 0.00521 \\
1.74 & 3.080 & 0.004 & 0.230 & 0.011 & -31.27 & 1.64 & 346.51 & 0.0173 & 0.00531 & 0.0068 & 0.00498 \\
1.83 & 3.030 & 0.005 & 0.235 & 0.012 & -30.06 & 1.70 & 383.25 & 0.0148 & 0.00563 & 0.00347 & 0.00533 \\
1.92 & 2.990 & 0.004 & 0.242 & 0.010 & -28.23 & 1.48 & 395.17 & 0.00763 & 0.00525 & 0.00344 & 0.00507 \\
2.01 & 2.960 & 0.004 & 0.249 & 0.009 & -26.52 & 1.27 & 442.78 & -0.00252 & 0.00528 & 0.00707 & 0.00516 \\
2.12 & 2.920 & 0.004 & 0.259 & 0.009 & -25.49 & 1.26 & 472.32 & -0.00662 & 0.00562 & 0.0114 & 0.00536 \\
2.22 & 2.890 & 0.004 & 0.273 & 0.009 & -25.32 & 1.19 & 507.07 & -0.00629 & 0.00546 & 0.0107 & 0.00523 \\
2.33 & 2.860 & 0.004 & 0.288 & 0.009 & -25.63 & 1.09 & 541.51 & -0.00244 & 0.00525 & 0.00999 & 0.00506 \\
2.45 & 2.820 & 0.004 & 0.299 & 0.008 & -26.44 & 0.94 & 581.27 & 0.00396 & 0.00511 & 0.00791 & 0.00496 \\
2.57 & 2.770 & 0.004 & 0.303 & 0.007 & -26.61 & 0.86 & 631.84 & 0.00946 & 0.0049 & 0.0116 & 0.00451 \\
2.70 & 2.710 & 0.004 & 0.301 & 0.008 & -26.39 & 0.91 & 697.4 & 0.0118 & 0.00519 & 0.0168 & 0.00455 \\
2.84 & 2.660 & 0.004 & 0.298 & 0.006 & -25.54 & 0.79 & 740.1 & 0.0101 & 0.00439 & 0.016 & 0.00382 \\
2.98 & 2.610 & 0.003 & 0.301 & 0.006 & -24.28 & 0.69 & 824.46 & 0.0102 & 0.00382 & 0.015 & 0.00328 \\
3.13 & 2.560 & 0.003 & 0.301 & 0.005 & -23.59 & 0.64 & 881.22 & 0.0124 & 0.00352 & 0.0117 & 0.00316 \\
3.28 & 2.490 & 0.003 & 0.301 & 0.004 & -23.46 & 0.53 & 961.84 & 0.0122 & 0.00297 & 0.00915 & 0.00265 \\
3.45 & 2.430 & 0.003 & 0.301 & 0.005 & -23.00 & 0.65 & 1050.4 & 0.0128 & 0.0039 & 0.00827 & 0.00358 \\
3.62 & 2.370 & 0.004 & 0.303 & 0.005 & -23.07 & 0.63 & 1127.2 & 0.0111 & 0.00378 & 0.00794 & 0.00349 \\
3.80 & 2.300 & 0.003 & 0.298 & 0.005 & -23.66 & 0.55 & 1230.3 & 0.00671 & 0.00332 & 0.00523 & 0.00316 \\
3.99 & 2.240 & 0.003 & 0.303 & 0.004 & -24.02 & 0.48 & 1331.5 & 0.00718 & 0.00297 & 0.00483 & 0.00285 \\
4.19 & 2.180 & 0.002 & 0.315 & 0.003 & -23.89 & 0.36 & 1411.8 & 0.00471 & 0.0024 & 0.00128 & 0.00224 \\
4.40 & 2.120 & 0.003 & 0.317 & 0.004 & -23.66 & 0.47 & 1523.8 & 0.00852 & 0.00289 & 0.00966 & 0.00249 \\
4.62 & 2.050 & 0.004 & 0.320 & 0.005 & -23.07 & 0.56 & 1644.2 & 0.00837 & 0.00356 & 0.011 & 0.00259 \\
4.85 & 1.970 & 0.004 & 0.316 & 0.004 & -22.44 & 0.52 & 1784.2 & 0.00626 & 0.00334 & 0.00866 & 0.00249 \\
5.09 & 1.890 & 0.004 & 0.312 & 0.004 & -22.05 & 0.53 & 1914.6 & 0.00111 & 0.00343 & 0.0114 & 0.0027 \\
5.35 & 1.800 & 0.004 & 0.304 & 0.004 & -22.18 & 0.55 & 2070.9 & 0.00185 & 0.00348 & 0.0168 & 0.00259 \\
5.61 & 1.730 & 0.003 & 0.303 & 0.004 & -21.98 & 0.49 & 2243.7 & 0.00247 & 0.00305 & 0.0163 & 0.00236 \\
5.89 & 1.650 & 0.003 & 0.302 & 0.004 & -21.50 & 0.52 & 2419.1 & 0.00633 & 0.00324 & 0.0174 & 0.0023 \\
6.19 & 1.580 & 0.003 & 0.302 & 0.004 & -21.78 & 0.50 & 2586.7 & 0.00358 & 0.00317 & 0.0173 & 0.00243 \\
6.50 & 1.490 & 0.003 & 0.294 & 0.004 & -21.06 & 0.55 & 2782.8 & 0.00264 & 0.00334 & 0.0188 & 0.0026 \\
6.82 & 1.420 & 0.003 & 0.286 & 0.004 & -19.33 & 0.54 & 2989.4 & -5.28E-4 & 0.00319 & 0.021 & 0.00235 \\
7.16 & 1.340 & 0.003 & 0.277 & 0.004 & -19.48 & 0.50 & 3226.4 & 0.00615 & 0.0028 & 0.0165 & 0.00234 \\
7.52 & 1.270 & 0.003 & 0.276 & 0.004 & -18.84 & 0.49 & 3446.2 & 0.0022 & 0.00277 & 0.0155 & 0.00238 \\
7.90 & 1.190 & 0.002 & 0.265 & 0.004 & -17.05 & 0.51 & 3710.3 & 0.00444 & 0.0027 & 0.0151 & 0.00244 \\
8.29 & 1.110 & 0.003 & 0.246 & 0.004 & -15.72 & 0.63 & 4006.3 & 1.00E-3 & 0.00312 & 0.0125 & 0.00295 \\
8.71 & 1.050 & 0.002 & 0.251 & 0.004 & -15.33 & 0.63 & 4232.9 & 0.00585 & 0.00324 & 0.0156 & 0.003 \\
9.14 & 0.980 & 0.002 & 0.243 & 0.005 & -15.33 & 0.69 & 4530.5 & 0.00633 & 0.0034 & 0.0113 & 0.00341 \\
\hline
\label{apptable3}\\
\end{longtable}
\end{center}

\end{appendix}

\end{document}